\documentclass[a4paper,11pt]{article}

\pdfoutput=1 

\usepackage{jheppub} 

\usepackage[T1]{fontenc} 

\title{\boldmath Non-split singularities and conifold transitions in F-theory}

\author[a]{R. Kuramochi,}
\author[a,b]{S. Mizoguchi
}
\author[c]{and T. Tani}

\affiliation[a]{SOKENDAI (The Graduate University for Advanced Studies)\\
Tsukuba, Ibaraki, 305-0801, Japan 
}
\affiliation[b]{Theory Center, 
Institute of Particle and Nuclear Studies,
KEK\\Tsukuba, Ibaraki, 305-0801, Japan }

\affiliation[c]{National Institute of Technology, Kurume College, \\
Kurume, Fukuoka, 830-8555, Japan}

\emailAdd{rinto@post.kek.jp}
\emailAdd{mizoguch@post.kek.jp}
\emailAdd{tani@kurume-nct.ac.jp}

\abstract{
In F-theory, if a fiber type of an elliptic fibration involves 
a condition that requires an exceptional curve to split 
into two irreducible components, it is called  ``split'' or ``non-split'' 
type depending on whether it is globally possible or not.
In the latter case, the gauge symmetry is reduced to a
non-simply-laced Lie algebra due to monodromy.
We show that this split/non-split transition is, 
except for a special class of models, 
a conifold transition from 
the resolved to the deformed side, associated with the conifold 
singularities emerging where the codimension-one singularity is 
enhanced to $D_{2k+2}$ $(k\geq 1)$ or $E_7$. 
We also examine how the previous proposal for
the origin of non-local matter can be actually 
implemented in our blow-up analysis.
}

\font\mybb=msbm10 at 12pt

\def\bb#1{\hbox{\mybb#1}}

\def\CC {\bb{C}}
\def\PP {\bb{P}}

\def\FF {\bb{F}}

\newcommand\beqa{\begin{eqnarray}}
\newcommand\eeqa{\end{eqnarray}}
\newcommand\n{\nonumber\\}
\begin{document}

\maketitle
\flushbottom

\section{Introduction}
In F-theory \cite{Vafa}, singularities play an essential role for the theory 
to geometrically realize various aspects of string theory \cite{MV1,MV2,BIKMSV,KatzVafa}. 
An F-theory compactified on an elliptic Calabi-Yau $n$-fold is 
basically a type IIB theory compactified on an $(n-1)$-dimensional 
base of a Calabi-Yau manifold with 7-branes in it, 
where the configuration of the axio-dilaton of type IIB string 
is described by that of the elliptic modulus of the fibration. 
A 
codimension-one locus of the base over which the elliptic fibers 
become singular
is the place where a collection of 7-branes reside on top of each other 
and typically realizes a non-abelian gauge 
symmetry depending on the fiber type 
following Kodaira's classification. Similarly, 
a  
codimension-two locus in the base is involved in matter generation.   
A codimension-three locus in the base is also possible in 
four-dimensional F-theory on a Calabi-Yau four-fold, involving 
Yukawa couplings.

Kodaira's classification \cite{Kodaira} of singular fibers of an elliptic surface 
is based on the intersection diagrams of exceptional curves 
that arise after the resolutions (table  \ref{tab:Kodaira}). 
For 
an elliptic Calabi-Yau $n$-fold which 
also allows a fibration of an elliptic surface over an $(n-2)$-fold, 
the singularities of 
(singular fibers of)
these fibered elliptic surfaces are 
aligned all the way along the $(n-2)$-fold, 
forming a codimension-two locus in the total elliptic Calabi-Yau $n$-fold, 
whose projection to the base (of the elliptic fibration) is 
the ``codimension-one'' locus mentioned above\footnote{We use here 
and below the scare quotes to emphasize that  
the codimension is counted in the base manifold of the 
elliptic fibration, and not in the total space. 
We do so because the use of such terminology 
was natural in the local F-GUT or the Higgs bundle approach 
especially popular in the late 00s and early 2010s 
(e.g. 
\cite{DonagiWijnholt,BHV,BHV2,DonagiWijnholt2,HKTW,DWHiggsBundles,
localmodel1,localmodel2,localmodel3,localmodel4}), 
but misleading when considering the geometry of the whole Calabi-Yau, including the fiber space.}.
We can blow up these ``codimension-one'' singularities 
in the base (codimension-two in the total space) to 
yield a collection of exceptional curves aligned along 
the ``codimension-one'' locus, 
so we can still talk about 
the fiber type of the singularity 
over a generic point on the ``codimension-one'' locus. 
\begin{table}[ht]
\caption{Kodaira's classification of singularities of an elliptic surface.}
\begin{center}
\begin{tabular}{r@{\hspace{12mm}}r@{\hspace{12mm}}c@{\hspace{10mm}}cc}
\hline 
ord$(f)$     &  ord$(g)$      &  ord$(\Delta)$    &   Fiber type  &$G$   \\  \hline
 $\geq 0$  &  $\geq 0$    &      $0$        &   smooth     &     none              \\ 
  $0$      &    $0$       &      $n$        &   $I_n$      &    $A_{n-1}$          \\
 $\geq 1$  &    $1$       &      $2$        &   $II$       &     none              \\
  $1$      &    $\geq 2$  &      $3$        &   $III$      &    $A_1$              \\
 $\geq 2$  &    $2$       &      $4$        &   $IV$       &    $A_2$              \\  
  $2$      &    $\geq 3$  &      $n+6$      &   $I_n^{*}$  &    $D_{n+4}$          \\
  $\geq 2$ &    $3$       &      $n+6$      &   $I_n^{*}$  &    $D_{n+4}$          \\
  $\geq 3$ &    $4$       &      $8$        &   $IV^{*}$   &    $E_6$              \\
  $3$      &    $\geq 5$  &      $9$        &   $III^{*}$  &    $E_7$              \\
 $\geq 4$  &    $5$       &      $10$       &   $II^{*}$   &    $E_8$              \\ 
 $\geq 4$  &    $\geq 6$  &   \hspace{-0.6cm}  $\geq 12$  &   non-minimal    &    \rule[0.3ex]{1ex}{0.05ex} \\ \hline
\end{tabular}   
\label{tab:Kodaira}
\end{center}
\end{table}
\begin{table}[h]
\caption{Singularities of the split and non-split types.
For the $I_{2k+1}$ fiber type, $I_{2k+1}^{os}$ denotes the ``over-split type''
which is explained in the text.
\label{split/nonsplit}
}
\begin{center}
\vskip -5ex
\begin{tabular}
{c@{\hspace{2ex}}c@{\hspace{2ex}}c@{\hspace{2ex}}c@{\hspace{2ex}}c@{\hspace{2ex}}c@{\hspace{-1ex}}c@{\hspace{-5ex}}c}
\hline 
 $
\begin{array}{c}
\raisebox{-1ex}{Kodaira's}\\
\raisebox{1ex}{fiber type}\\
\end{array}
$&
ord$(b_2)$&  ord$(b_4)$      &  ord$(b_6)$    & ord$(b_8)$  &ord$(\Delta)$&
$
\begin{array}{c}
\raisebox{-1ex}{Additional}\\
\raisebox{1ex}{constraint(s)}\\
\end{array}
$
&
 $
\begin{array}{c}
\raisebox{-1ex}{Split/non-split}\\
\raisebox{1ex}{fiber type}\\
\end{array}
$
    \\  \hline
$I_{2k}(k\geq 2)$ &  $0$ & $k$ &  $2k$ & $2k$ & $2k$ & 
$b_{2,0}=
c_{1,0}^2$& $I_{2k}^s$          \\
&&&&&&~~$b_{2,0}$~\mbox{generic}&$I_{2k}^{ns}$ 
\\  \hline
$I_{2k+1}(k\geq 1)$ &  $0$ & $k$ &  $2k$ & $2k+1$ & $2k+1$ & 
$
\left\{\begin{array}{l}
b_{2,0}= c_{1,0}^2\\
b_{4,k}= c_{1,0}c_{3,k}\\
b_{6,2k}=c_{3,k}^2
\end{array}
\right.
$
& $I_{2k+1}^s$          \\
&&&&&&
$
\left\{\begin{array}{l}
b_{2,0}~\mbox{generic}\\
b_{4,k}= b_{2,0}c_{2,k}\\
b_{6,2k}=b_{2,0}c_{2,k}^2
\end{array}
\right.
$
&$I_{2k+1}^{ns}$ 
 \\
&&&&&&
$
\left\{\begin{array}{l}
b_{2,0}= c_{1,0}^2\\
b_{4,k}= c_{1,0}^2 c_{2,k}\\
b_{6,2k}=c_{1,0}^2 c_{2,k}^2
\end{array}
\right.
$
&$I_{2k+1}^{os}$ 
\\  \hline
$I_{0}^*$ &  $1$ & $2$ &  $3$ & $4$ & $6$ & 
$
\left\{\begin{array}{l}
b_{2,1}=4(p_{2,1}+q_{2,1}+r_{2,1})\\
b_{4,2}=2(p_{2,1}q_{2,1}+q_{2,1}r_{2,1}+r_{2,1}p_{2,1}) \\
b_{6,3}=4p_{2,1}q_{2,1}r_{2,1}
\end{array}
\right.
$
&$I_{0}^{*s}$         \\
&&&&&&~$
\left\{\begin{array}{l}
b_{2,1}=4(p_{2,1}+q_{2,1})\\
b_{4,2}=2(p_{2,1}q_{2,1}+r_{4,2}) \\
b_{6,3}=4p_{2,1}r_{4,2}
\end{array}
\right.
$
&$I_{0}^{*ss}$  \\
&&&&&&$
\begin{array}{c}
\raisebox{-1ex}{$b_{2,1},b_{4,2},b_{6,3}$}\\
\raisebox{1ex}{generic}
\end{array}
$
&$I_{0}^{*ns}$          
\\  \hline
$I_{2k-3}^*\,(k\geq 2)$ &  $1$ & $k+1$ &  $2k$ & $2k+1$ & $2k+3$ & 
$b_{6,2k}=
c_{3,k}^2$& $I_{2k-3}^{*s}$          \\
&&&&&&~~$b_{6,2k}$~\mbox{generic}&$I_{2k-3}^{*ns}$ 
\\  \hline
$I_{2k-2}^*\,(k\geq 2)$

&  $1$ & $k+1$ &  $2k+1$ & $2k+2$ & $2k+4$ & 
$b_{8,2k+2}=
c_{4,k+1}^2$& $I_{2k-2}^{*s}$          \\
&&&&&&~~$b_{8,2k+2}$~\mbox{generic}&$I_{2k-2}^{*ns}$ 
\\  \hline
$IV$ &  $1$ & $2$ &  $2$ & $3$ & $4$ & 
$b_{6,2}=
c_{3,1}^2$& $IV^{s}$          \\
&&&&&&~~$b_{6,2}$~\mbox{generic}&$IV^{ns}$ 
\\  \hline
$IV^*$ &  $2$ & $3$ &  $4$ & $6$ & $8$ & 
$b_{6,4}=
c_{3,2}^2$& $IV^{*s}$          \\
&&&&&&~~$b_{6,4}$~\mbox{generic}&$IV^{*ns}$ 
\\  \hline
  \end{tabular}   
\label{tab:Splitnonsplit}
\end{center}
\end{table}

In these lower-dimensional 
F-theories, 
unlike the eight-dimensional theory 
on just a single elliptic surface, 
if the fiber type involves 
a condition that requires an exceptional curve to split 
into two irreducible components, 
these two split curves on a generic point generally
meet on top of each other at some point along the ``codimension-one'' locus.
If such exceptional fibers 
of an elliptic surface 
constitute part of the same smooth irreducible locus in the total space of the
Calabi-Yau, the fiber type is called ``non-split'' \cite{BIKMSV}.
If this happens, 
the two apparently distinct exceptional fibers 
are swapped with each other at some point 
when one goes along the $(n-2)$-fold, and 
hence are considered to be identical. This phenomenon is known as
a monodromy.
The expected $G$ (simply-laced) gauge symmetry is then 
subject to a 
projection by a diagram automorphism,  
reduced to a corresponding non-simply-laced gauge symmetry.
Such identification of exceptional 
fibers can occur when the fiber type is $I_{n}$ $(n=3,4,\ldots)$, 
$I_n^*$ $(n=0,1,\ldots)$, $IV$ or $IV^*$.
If, on the other hand, the two split exceptional fibers of 
each elliptic surface belong to different irreducible  exceptional surfaces 
in the total Calabi-Yau and hence are split globally, 
the fiber type is called ``split'', yielding the expected 
$G$ gauge symmetry implied by Kodaira's classification  \cite{BIKMSV}.

The points where the two exceptional curves overlap 
constitute a special codimension-two locus in the base space
(of the elliptic fibration), 
where the singularity is enhanced from $G$ to higher  
\footnote
{
As is well known and shown in table  \ref{tab:Kodaira}, 
there is an almost one-to-one  
correspondence between a Kodaira fiber type and 
a   Dynkin diagram of some simply-laced Lie algebra $G$ 
(except for $G=SU(2)$ and $SU(3)$), 
so we may say  ``the  singularity is $G$'' by using the corresponding 
Lie  algebra. 
Note that, as is also known, the intersection diagram 
deduced from the apparent Kodaira fiber type 
found fiber-wise may or may not coincide with the intersection 
diagram of the actual exceptional curves that emerge
through the blow-ups performed to resolve the singularities. 
}
in the sense of the fiber type of Kodaira directly over 
that point.
In the split case, 
there typically (but not always) arises a conifold 
singularity \cite{MT,EsoleYau}, and a wrapped M2-brane  
(in the M-theory dual) around a new two-cycle, 
which emerges due to the small resolution, 
accounts for the generation of the localized 
matter multiplet \cite{KatzVafa}.

For example, in a six-dimensional F-theory with $SU(5)$ 
gauge symmetry compactified on an elliptic Calabi-Yau 
3-fold over a Hirzebruch surface $\FF_n$, 
 there are $n+2$ codimension-two loci on the base 
 where a generic split $I_5$ fiber becomes 
 $I_1^*$, and $3n+16$ loci where it becomes 
 $I_6$\footnote{Strictly speaking, this is the case when the ``gauge divisor''
(the divisor representing the stack of 7-branes in IIB theory 
carrying a nonabelian gauge symmetry) is taken to be 
$D_u=D_v+n D_{u'}$ 
in the notation of section 2. One can alternatively take $D_v$, but then, since 
$D_v^2=-n<0$ and $KD_v=n-2$, where $K$ is the canonical divisor, $-4K-D_v$
becomes effective if $n>2$. This means that the section $f$ vanishes
on the divisor $D_v$, 
therefore the fiber type cannot be $I_5$ in this case \cite{MT1201.1943}.}. 
%
Therefore, if a {\bf 10} of $SU(5)$ appears at the ``$SO(10)$ point'' 
and {\bf 5} at the ``$SU(6)$ point''\footnote{
In the following, we will refer to a point on the base 
of the ${\rm dP}_9$ fibration as a ``$G$ point'' if Kodaira's classification 
of the singular fiber just over that point corresponds 
to a Lie algebra $G$.
},
 they 
together with the $5n+36$ neutral hypers from the complex structure moduli 
exactly satisfy the  anomaly cancellation condition $n_H-n_V=30n+112$ 
\cite{GSWest,BIKMSV}. 

On the other hand, in the non-split $I_5$ case, 
while {\bf 5}'s are still expected to appear at the 
$SU(6)$ points 
where the structure of the singularity does not change, 
the anomaly cancellation condition cannot be satisfied 
no matter what kind of matter field is assumed to be 
{\em locally} generated 
at the 
$SO(10)$ points, which are twice as many as the split case.
On top of that, the conifold singularity does not appear, even 
though the singularity 
in the sense of Kodaira is apparently enhanced 
to $SO(10)$ over that point. 
Rather, by blowing up a nearby ``codimension-one'' singularity, 
the singularity there is simultaneously resolved together.
Thus there is no sign of a localized matter field in the non-split case, 
although the anomaly cancellation condition (in six dimensions in particular)
requires a definite amount of chiral matter field to arise 
 even in the non-split model with a non-simply-laced 
gauge symmetry.
%
Such a phenomenon is widespread in other non-split models 
\cite{BIKMSV,AKM,GHLST,AGW,EJK,EK,EJ}
\footnote{
We note that 
analogous phenomena occur on codimension-two loci in 6D split models, 
where the gauge symmetries are $SU(6)$, $SO(12)$ and $E_7$,   
and the singularities are enhanced from those to $E_6$, $E_7$ and $E_8$, 
respectively \cite{MT, KMT,KuMT}.
The matter multiplets generated there are 
half-hypermultiplets in some pseudo-real representations.
In this case as well, the conifold singularity does not arise, but 
here the intersection of the exceptional curves  changes there
so that a root corresponding to an exceptional curve splits off 
into two (or more) weights \cite{Yukawas}. 
Such a matter field generation 
was called an ``incomplete resolution'' \cite{MT}.
}.

In fact, Ref.~\cite{AKM} has proposed 
a mechanism for non-local matter generation 
 that does not require any new exceptional curves 
 in those non-split models.
 The idea is as follows: As mentioned above, 
 in a non-split model, some of the exceptional curves 
are identified in pairs. Each such pair forms a ruled (=$\PP^1$-fibered)
(complex) surface whose base is a (real two-dimensional) 
Riemann surface of genus $g$, where $2g+2$ is the number 
of the places where the conifold singularities 
disappear on the gauge divisor in the transition to the non-split model.
According to Witten's and Katz-Morrison-Plesser(KMP)'s 
discussions \cite{Witten,KMP}, they claim that from each pair 
there appear $g$ 
hypermultiplets from (the harmonic 1-forms of) the genus-$g$ Riemann surface, 
corresponding to a {\em short} simple root of the non-simply-laced 
gauge Lie algebra of the non-split model. It was argued that all these 
hypermultiplets, together with ones coming from the monodromy-invariant 
exceptional curves, generate the whole desired representation at the 
desired multiplicities  \cite{AKM}.
 
This proposal works well, but some questions remain:\\
(1) In split models, conifold singularities appear, and hypermultiplets 
are generated from their small resolutions.
On the other hand, \cite{AKM} proposed that hypermultiplets 
appear from the genus-$g$ Riemann surface.
Are they related, and if so, how?\\
(2) \cite{AKM} is based on Witten's and KMP's discussions 
of hypermultiplet generation, but in the latter 
the situation is that there is originally a genus-$g$ Riemann surface 
with a singularity on it, whereas in the present case the gauge divisor is genus-zero.
(That is consistent with the spectrum not including the adjoint hyper.)
Certainly a blow-up gives rise to a ruled surface on a genus-$g$ 
Riemann surface, but if there are multiple pairs of exceptional curves 
that are identified by monodromy, it appears that multiple ruled surfaces 
on different genus-$g$ Riemann surfaces arise. 
Can we still in that case apply \cite{AKM}'s argument 
which assumes only a single  Riemann surface of genus $g$?

In this paper, 
we perform blow-ups in all models 
where there is a distinction between split and non-split 
fiber types, and by doing so we specifically examine the proposal of \cite{AKM} for non-local generation of matter fields in the non-split 
models of F-theory
\footnote{
Various different patterns of the intersections of the exceptional curves 
arising from different singularity resolutions of the split models were 
studied in  \cite{1402.2653} by means of 
the Coulomb branch analysis of the M-theory gauge theory.
}.
We will deal with 
a 6D F-theory compactified on 
an elliptic Calabi-Yau threefold (and stable degenerations thereof) 
on the Hirzebruch surface 
\cite{MV1,MV2,BIKMSV}. 
%
We will show that the split/non-split transition is, 
except for some special cases, 
a conifold transition 
\cite{CdlOGP,CGH,CdlO,CdlO2,AGM,Strominger} 
from 
the resolved to the deformed side, associated with conifold 
singularities emerging at codimension-two loci in the base 
of the split models. 
%
The ``genus-$g$ curve'' of \cite{AKM} can be obtained 
as an intersection of this local deformed conifold and 
some appropriate divisors. This will answer the question (1).

%
It turns out that there are several different patterns of the transition.
We found that,
at the codimension-two 
loci of the base where the relevant conifold 
singularities arise after the codimension-one blow-ups,  
 the singularity there is always enhanced to $D_{2k+2}$ ($k\geq1$), or $E_7$ which is the only case for 
the $IV^{*s}\leftrightarrow IV^{*ns}$ transition, 
except 
for a special class of models in which 
no conifold singularity appears 
at the relevant codimension-two 
singularity in the split models.
%
We will also show that the $I_{2k+1}$ split models ($k\geq 1$),
which do not generically have the $D_{2k+2}$ enhanced points,
do not transition directly to $I_{2k+1}$ non-split models, 
but do via special $I_{2k+1}$ split models where 
the complex structure is tuned in such a way that they may develop  
$D_{2k+2}$ enhanced points. We will refer to these specially tuned 
split models as ``over-split'' models and denote them by $I_{2k+1}^{os}$.

 In all these cases where a conifold transition occurs, the 
 conifold singularities are resolved in the split models 
 by small resolutions to yield exceptional curves which are 
 two-cycles. Thus the split models are on the resolved side 
 of the conifold transition. On the other hand, we will show that 
 modifying the sections relevant to the transition from the split to the non-split 
 model amounts to deforming the conifold singularities to yield 
 local deformed conifolds, where three-cycles 
 appear instead of two-cycles on the resolved (split) side.
 
 We will explicitly show the equation of the genus-$g$ curve  
 for each pair of exceptional curves that join smoothly at a branch point 
 of it represented as a 2-sheeted Riemann surface. 
 Since such a complex two-surface 
 ``swept'' by a pair of curves constitutes a root of 
 the non-simply-laced gauge Lie algebra and arises at each step 
 of the blow-up, a priori the genus-$g$ base could be different 
 for each pair if there is more 
 than one pair of exceptional curves that are identified (in the 
 cases of $C_k$ and $F_4$).
 If so, that would be a problem, since \cite{AKM} assumes 
 the existence of only one genus-$g$ base (``$M_2$'' 
 in \cite{AKM}),   
 whose zero modes are responsible for the generation of 
 the necessary hypermultiplets. Happily, we show that  
 the genus-$g$ bases are all identical and common in these cases, 
 so we can have a well-defined single genus-$g$ base for these 
 cases as well.
 This is the answer to the question (2).
 


The rest of this paper is organized as follows.
In section 2, we summarize the basic set-ups of 
6D F-theory on an elliptic Calabi-Yau threefold 
over a Hirzebruch surface, which will be used in the 
subsequent sections.
Sections 3 to 7 consider separately all fiber types 
in which a non-split type exists.
We deal with the $I_{2k}$ models first in section 3, 
and show in detail that the split/non-split transition there 
is a conifold transition associated with the conifold 
singularities arising at the $D_{2k}$ points. 
The last subsection 
gives a short summary of the proposal of \cite{AKM} for 
non-local matter generation and answer to the questions 
stated above.
The fiber types $I_{2k+1}$, $IV$ and $IV^*$ are 
studied in sections 4,5 and 6, respectively, and a 
similar conclusion is reached, with an exception that 
the relevant conifold transition occurs at the $E_7$ points 
in the $IV^*$ model. Section 7 is devoted to the study of 
the final example, the $I_n^*$ models, where it is shown 
that in the $I_{2k-3}^*$ models, the split/non-split transition 
is similarly understood as a conifold transition, while 
in the $I_{2k-2}^*$ models, no conifold singularity 
arises at the relevant enhanced points.
Finally, we conclude in section 8.

\section{Summary of 6D F-theory on an elliptic Calabi-Yau threefold 
over a Hirzebruch 
surface}

Let us consider six-dimensional F-theory compactified on an elliptic 
Calabi-Yau threefold  $Y_3$ with a section fibered over a Hirzebruch surface 
$\FF_n$ 
($n\geq 0$)
\cite{MV1,MV2}.
We define $Y_3$ as a 
hypersurface 
\beqa
&&-(y^2+a_1 x y + a_3 y)+x^3 + a_2 x^2 + a_4 x + a_6 =0
\label{Tate's}
\eeqa
in a complex four-dimensional 
ambient space $X_4$, which itself is 
a $\PP^2$ fibration over $\FF_n$.
$(x, y)$ are the affine coordinates in a coordinate patch of $\PP^2$
where one of the homogeneous coordinates does not vanish 
and hence is set to 1.
Let ${\cal K}$ be the canonical bundle of $\FF_n$, then 
$x$, $y$ are sections of ${\cal K}^{-2}$, ${\cal K}^{-3}$,
whereas $a_j$ $(j=1,2,3,4,6)$ are ones of ${\cal K}^{-j}$, respectively, 
so that the hypersurface (\ref{Tate's}) defines a Calabi-Yau threefold. 

A Hirzebruch surface $\FF_n$ is a $\PP^1$ fibration over $\PP^1$,
defined as a toric variety 
with the following toric charges
\beqa
\begin{tabular}{ccccl}
          &~$u'$   &~$v'$   &~$u$  &~$v$\\
$Q^1$&~$1$   &~$1$     &~$n$  &~$0$\\
$Q^2$&~$0$   &~$0$     &~$1$  &~$1$~~.\\
\end{tabular}
\eeqa
$(u':v')$ are the homogeneous coordinates of the base $\PP^1$,
while $(u:v)$ are the ones of the fiber $\PP^1$.
The anti-canonical 
bundle  
corresponds to 
the divisor $(n+2)D_{u'}+2D_v$, where we denote, for 
a given coordinate $X$, 
by $D_X$ a divisor 
defined by the zero locus $X=0$. 
Thus, if we define affine coordinates 
$z\equiv \frac uv$, $w\equiv\frac{u'}{v'}$ in a patch 
$v\neq 0$ and $v'\neq 0$, the section 
$a_j$ is given as a $2j$th degree polynomial in $z$ and 
a $j(n+2)$th degree polynomial in $w$.

The hypersurface so defined is also a K3 fibration, the base of which is  
the base $\PP^1$ of $\FF_n$.
We next consider the stable degeneration limit of this K3.
Schematically, this is regarded as a limit of splitting into a
pair of rational elliptic surfaces ${\rm dP}_9$ glued together along 
the torus fiber over the ``infinite points'' of the respective bases.
See \cite{MV2,Aspinwall} for a more rigorous definition.

It is convenient to move on to a 
${\rm dP}_9$ fibration over the same  
$\PP^1$ with $u',v'$ being its coordinates.
To do this, we have only to change the divisor class of 
$a_j$ from 
$j((n+2)D_{u'}+2D_v)$ ($=$ the divisor of ${\cal K}^{-j}$) to 
\beqa
j((n+2)D_{u'}+D_v). \label{ajdivisor}
\eeqa
With this change, 
$a_j$ is still 
a $j(n+2)$th degree polynomial in $w$ but becomes $j$th degree in $z$.
Likewise, the divisor classes of $x$ and $y$ are modified from 
 %
$2((n+2)D_{u'}+2D_v)$,  
$3((n+2)D_{u'}+2D_v)$
to
\beqa
2((n+2)D_{u'}+D_v),  ~~3((n+2)D_{u'}+D_v),
\label{xydivisor}
\eeqa
respectively. 
This 
${\rm dP}_9$ fiber describes one $E_8$ of the
 $E_8 \times E_8$ gauge symmetry. 
The terms of degrees from $j+1$ to $2j$ appearing 
in $a_j$ for the K3 fibration correspond to the other  
${\rm dP}_9$ residing ``beyond the infinity''. 
For generic ${\rm dP}_9$ fibrations,  
$a_j$ is expanded as
\beqa
a_j=a_{j,0}+a_{j,1}z+\cdots+a_{j,j-1}z^{j-1}+a_{j,j}z^j~~~(j=1,2,3,4,6),
\eeqa
then the section $a_{j,k}$ of each coefficient becomes a 
($(j-k)n+2j$)th degree polynomial in $w$
due to the nonzero $Q^1$ charge carried by $u$.

As an equation of an elliptic fiber, 
(\ref{Tate's}) is commonly referred to as  ``Tate's form''.
One can complete the square with respect to $y$ in (\ref{Tate's}) 
to obtain (with a redefinition of $y$)
\beqa
&&-y^2+x^3 +\frac{b_2}4 x^2 + \frac{b_4}2 x + \frac{b_6}4 =0,\label{Deligne}
\eeqa
\beqa
b_2&=&a_1^2+4a_2,\n
b_4&=&a_1 a_3+2 a_4,\\
b_6&=&a_3^2+4 a_6,\nonumber
\eeqa
which, though less common, we call the ``Deligne form'' in this paper 
\cite{Deligne}.
$b_j$ is a section of the same line bundle as $a_j$ and similarly expanded as 
\beqa
b_j=b_{j,0}+b_{j,1}z+\cdots+b_{j,j-1}z^{j-1}+b_{j,j}z^j~~~(j=2,4,6),
\eeqa
where $b_{j,k}$ is also  a ($(j-k)n+2j$)th degree polynomial in $w$.
It is also convenient to define \cite{BIKMSV} 
\beqa
b_8&=&\frac14(b_2 b_6- b_4^2),\label{b8}
\eeqa
which is the (minus of the) 
discriminant of the quadratic equation 
\beqa
\frac{b_2}4 x^2 + \frac{b_4}2 x + \frac{b_6}4 =0
\eeqa
of $x$.

Finally, one can ``complete the cube'' with respect to $x$ in (\ref{Deligne})
and find (with a redefinition of $x$)
 \beqa
&&-y^2+x^3 +f 
x 
+ g =0,\label{Weierstrass}
\eeqa
\beqa
f&=&-\frac{1}{48}(b_2^2-24 b_4),\n
g&=&\frac{1}{864} \left(b_2^3-36 b_2 b_4+216 b_6\right),
\label{fandg}
\eeqa
which is called the ``Weierstrass form''.  $f$ and $g$ are 
sections of the same line bundle as $a_4$ and $a_6$, respectively, 
and in the ${\rm dP} _9$ 
fibration they are expanded as 
\beqa
f&=&f_{4,0}+f_{4,1}z+\cdots+f_{4,4}z^4,\n
g&=&g_{6,0}+g_{6,1}z+\cdots+g_{6,6}z^6,
\eeqa
where $f_{4,k}$,  $g_{6,k}$ are written as $f_{(4-k)n+8}$, $g_{(6-k)n+12}$
in \cite{MV1}, 
whose degrees in $w$ are specified by their subscripts.
The discriminant $\Delta$ of (\ref{Weierstrass}) is 
\beqa
\Delta&=&4f^3+27 g^2\n
&=&\frac{1}{16} \left(b_2^2 b_8 -9 b_2 b_4
   b_6+8 b_4^3+27 b_6^2\right).
\eeqa

Consider the case where the elliptic fiber over $z=0$ of the base 
$\PP^1$ of this $ {\rm dP} _9 $ (i.e. the fiber $\PP^1$ of the $\FF_n$) 
has a singularity, and the exceptional fibers after the resolution fall into 
one of Kodaira's fiber types.
It is well-known that 
the fiber type of a given singularity is 
determined in terms of the vanishing orders of the sections 
$f$, $g$ of the Weierstrass form as well as the discriminant 
$\Delta$ (table  \ref{tab:Kodaira}).

Note that, in Kodaira's classification, 
there is no upper limit on the vanishing orders of $f$, 
$g$ or $\Delta$
(since any large value of  $n$ is allowed for the 
fiber type $I_n$ or $I^*_n$ as a fiber type)\footnote{Of course, as is well known, 
if the orders of f and g increase simultaneously to 4 and 6, 
the resulting singularities will have bad properties.}
, but there {\em is}  
when we try to realize singular fibers in a ${\rm dP}_9$ fibration.
Since the relationship between the split/non-split transition 
and the conifold transition discussed below is also a local one 
in the sense that it does not depend on another 
singularity located far away, we will also need 
to consider a high vanishing 
order that cannot be realized in a ${\rm dP}_9$ fibration. 
So in this paper, we will first start from a ${\rm dP}_9$ fibration 
and consider heterotic duality 
when it makes sense, 
while discussing the relationship  between the two transitions locally
in the same set-up even when the fiber cannot be realized 
in a ${\rm dP}_9$ fibration. 
 


As we already described in Introduction, if the type of 
a singular fiber is either $I_{n}$ $(n=3,4,\ldots)$, 
$I_n^*$ $(n=0,1,\ldots)$, $IV$ or $IV^*$ 
at a generic point $w$ on the divisor $z=0$ in $\FF_n$,
it is further classified as a split type or a non-split type, 
depending on whether or not the split condition 
is satisfied globally 
\footnote{$k=1$ ($I_0^*$) is a special case because 
there are three different types (split, non-split and semi-split) in this case; 
see \cite{EsoleetalSO(8)} for details.} . 
We have listed them in table  \ref{split/nonsplit}  
together with the required 
constraints for the fibers to be classified into 
the respective types
\footnote{Note that the vanishing orders for $b_i$'s ($i=2,4,6,8$)
presented here are, unlike the conventional orders in Tate's form 
\cite{BIKMSV,AKM,KMSNS},
the ones which are such that a given fiber type can 
be described by {\em generic} $b_i$'s with these orders.
For example, the orders of the sections $a_i$'s determining Tate's form 
 ($i=1,2,3,4,6$) for the non-split $I_{2k+1}$ model are
known to be $(0,0,k+1,k+1,2k+1)$, which imply the orders 
of $b_4$ and $b_6$ calculated using these data are 
$k+1$ and $2k+1$ instead of $k$ and $2k$.
These Tate's orders are the ones that are maximally raised 
within what a given fiber type can achieve, and only the specially 
tuned sections with appropriate redefinitions of $x$ and $y$ can
satisfy the condition. Indeed, as we show explicitly below, 
the orders of the generic $b_4$ and $b_6$ 
that can achieve a non-split $I_{2k+1}$ model 
are $k$ and $2k$.
}.
In the following, we will study these individual cases.

\section{Split/non-split transitions as conifold transitions (I): the $I_{2k}$ 
models}

\subsection{Generalities of the $I_n$ models}
\label{subsection:generalities}
Let us first summarize the generalities of the $I_n$ models 
common to both cases when $n$ is even and when $n$ is odd
\footnote{The resolutions of the split $I_n$ and $I_n^*$ models for 
even and odd $n$ were already computed in detail in \cite{1212.2949}. }.
As displayed in table  \ref{split/nonsplit}, 
the vanishing orders of the sections $b_2$, $b_4$ and $b_6$ of  (\ref{Deligne}) 
are $(0,k,2k)$ for both $I_{2k}$ and $I_{2k+1}$. 
The only difference is that the order of $b_8$ (\ref{b8}) 
is the generic value $2k$ in the $I_{2k}$ type, 
while in the $I_{2k+1}$ type $b_2$, $b_4$ and $b_6$ 
take special values so that the order of $b_8$ goes up to $2k+1$.
Explicitly, the equation of these models is given by
\beqa
\Phi(x,y,z,w)\equiv&-y^2+x^3 &+{\textstyle \frac14} (b_{2,0}+b_{2,1}z+\cdots) x^2 \n
&&
+ {\textstyle \frac12}  (b_{4,k}z^k+b_{4,k+1}z^{k+1}+\cdots)x 
\n&&
+{\textstyle \frac14} (b_{6,2k}z^{2k}+b_{6,2k+1}z^{2k+1}+\cdots)=0.
\label{I_2kI_2k+1}
\eeqa
As mentioned at the end of the previous section, this equation is not well defined as a ${\rm dP}_9$ fibration when $k$ is large (e.g., $k\geq 4$), 
but even in that case we will use it to analyze the local structure near the conifold singularities 
associated with the split/non-split transition.

The equation (\ref{I_2kI_2k+1})
has a singularity at $(x,y,z)=(0,0,0)$ for arbitrary $w$ in both cases. 
We will blow up this singularity, as well as the ones we will subsequently encounter, 
by taking the usual steps. 
Let us explain the general procedure of how this is done by 
taking the present case as an example.
Our notation is similar to the one used in our previous paper \cite{KMT}.

We first replace the point $(x,y,z)=(0,0,0)$ in the complex three-dimensional 
$(x,y,z)$ space, which is a local patch of the three-dimensional 
ambient space defining the ${\rm dP}_9$,  by a $\PP^2$ 
by replacing
$\CC^3\ni(x,y,z)$ with 
\beqa
\hat{\CC}^3&=&\{
((x,y,z),(\xi:\eta:\zeta))\in\CC^3\times\PP^2
~|~(x:y:z)=(\xi:\eta:\zeta)
\}.
\label{C3hat}
\eeqa
We work in inhomogeneous coordinates defined in
three different patches of this $\PP^2$ 
\beqa
(x:y:z)=(\xi:\eta:\zeta)&=&(1:y_1:z_1)~~~\mbox{(${\bf 1}_x$, $x\neq 0$),}\n
&=&(x_1:1:z_1)~~~\mbox{(${\bf 1}_y$, $y\neq 0$),}\n
&=&(x_1:y_1:1)~~~\mbox{(${\bf 1}_z$, $z\neq 0$)},
\label{patches1}
\eeqa
where ${\bf 1}_x$, ${\bf 1}_y$ and ${\bf 1}_z$ are the names of the coordinate 
patches.\footnote{In (\ref{patches1}), one and the same symbol represents 
two different variables in different equations ($y_1$ in ${\bf 1}_x$ and ${\bf 1}_z$, 
for instance). There will be no confusion, however, since these two patches will not be considered at the same time.}
Then replacing $\CC^3$ with $\hat{\CC}^3$ 
(\ref{C3hat}) is simply achieved by replacing $(x,y,z)$ with 
$(x, x y_1, x z_1)$ in  ${\bf 1}_x$, 
$(x_1 y, y, y z_1)$ in  ${\bf 1}_y$ and  
$(x_1 z, y_1 z, z)$ in  ${\bf 1}_z$ in the equation (\ref{I_2kI_2k+1}), respectively,
followed by dividing by the square of the scale factor
\beqa
 x^{-2}\Phi(x,x y_1,x z_1,w)\equiv\Phi_x(x,y_1,z_1,w)=0~~~\mbox{(${\bf 1}_x$),}\n
 y^{-2}\Phi(x_1 y, y, y z_1,w)\equiv\Phi_y(x_1, y, z_1,w)=0~~~\mbox{(${\bf 1}_y$),}\n
z^{-2}\Phi(x_1 z, y_1 z, z,w)\equiv\Phi_z(x_1, y_1, z,w)=0~~~\mbox{(${\bf 1}_z$)}
\label{In1stblowup}
\eeqa
so as not to change the canonical class.

Then we see that, unless $k=1$ ($I_2$ and $I_3$),
another singularity appears in the patch ${\bf 1}_z$ at $(x_1, y_1, z)=(0,0,0)$, 
then we do a similar replacement and factorization
\beqa
 x_1^{-2}\Phi_z(x_1,x_1 y_2,x_1 z_2,w)\equiv\Phi_{zx}(x_1,y_2,z_2,w)=0~~~\mbox{(${\bf 2}_{zx}$),}\n
 y_1^{-2}\Phi_z(x_2 y_1, y_1, y_1 z_2,w)\equiv\Phi_{zy}(x_2, y_1, z_2,w)=0~~~\mbox{(${\bf 2}_{zy}$),}\n
z^{-2}\Phi_z(x_2 z, y_2 z, z,w)\equiv\Phi_{zz}(x_2, y_2, z,w)=0~~~\mbox{(${\bf 2}_{zz}$).}
\label{In2ndblowup}
\eeqa
for each patch of another $\PP^2$ put at $(x_1, y_1, z)=(0,0,0)$.
Again, if $k$ is larger than two, we find a singularity in the patch 
${\bf 2}_{zz}$, which we blow up to obtain $\Phi_{zzz}(x_3,y_3,z,w)$. 
Repeating these steps $k$ times yields 
$\Phi_{\underbrace{\mbox{\scriptsize $z\cdots z$}}_k}(x_k,y_k,z,w)$, the properties of which 
differ between the types $I_{2k}$ and $I_{2k+1}$.

In the following, we will use the following $j$-times blown-up equations
recursively defined 
by 
\beqa
z^{-2}
\Phi_{\underbrace{\mbox{\scriptsize $z\cdots z$}}_{j-1}}(x_j z, y_j z, z,w)
&\equiv&
\Phi_{\underbrace{\mbox{\scriptsize $z\cdots z$}}_{j}}(x_j, y_j, z,w)
=0~~~
(\mbox{\boldmath $j$}_{\underbrace{\mbox{\scriptsize $z\cdots z$}}_{j}}),
\label{Injthblowup}
\\
x_{j-1}^{-2}
\Phi_{\underbrace{\mbox{\scriptsize $z\cdots z$}}_{j-1}}(x_{j-1}, x_{j-1}y_j, x_{j-1}z_j,w)
&\equiv&
\Phi_{\underbrace{\mbox{\scriptsize $z\cdots z$}}_{j-1}x}(x_{j-1}, y_j, z_j,w)
=0~~~
(\mbox{\boldmath $j$}_{\underbrace{\mbox{\scriptsize $z\cdots z$}}_{j-1}x})
\label{Injthblowupzx}
\eeqa
from the $(j-1)$-times blown-up equation
$\Phi_{\underbrace{\mbox{\scriptsize $z\cdots z$}}_{j-1}}(x_{j-1}, y_{j-1}, z,w)
=0$
defined in 
the coordinate patch 
$\mbox{\boldmath $(j\!-\!1)$}_{\underbrace{\mbox{\scriptsize $z\cdots z$}}_{j-1}}$. 
(Again, $y_j$'s in (\ref{Injthblowup}) and  (\ref{Injthblowupzx}) 
are different.)

\subsection{``Codimension-one'' singularities of the $I_{n}$ models}
We have seen in the previous subsection that there appears a singularity 
in ${\bf 1}_z$ at $(x_1, y_1, z)=(0,0,0)$ for arbitrary $w$, and after the blow-up 
there is, if $k\geq 3$,  another 
at $(x_2, y_2, z)=(0,0,0)$ in ${\bf 2}_{zz}$ for arbitrary $w$. These singular ``points'' 
in the sense of Kodaira are aligned along the base $\PP^1$ of $\FF_n$,
and hence form complex one-dimensional curves. If, though not considered  
in this paper, our set-up is generalized to a 4D F-theory compactification 
where the ${\rm dP}_9$ is fibered on some complex two-dimensional 
base, these singularities are aligned to form complex surfaces. Thus, 
in this paper, we will call such a singularity in the sense of Kodaira, 
that forms a codimension-one locus when projected onto the base 
of the elliptic fibration, a ``codimension-one''  singularity.

Using this terminology, we can say that, in the process of blowing up,
both the $I_{2k}$ and $I_{2k+1}$ models yield 
a ``codimension-one'' singularity $p_j$ at 
$(x_j, y_j, z,w)=(0,0,0,w)$ for every $j=0,\ldots,k-1$ in 
${\mbox{\boldmath $j$}}_{\underbrace{\mbox{\scriptsize $z\cdots z$}}_{j}}$, 
where we define  $(x_0, y_0, z,w)\equiv(x,y,z,w)$.
The explicit form of $\Phi_{\underbrace{\mbox{\scriptsize $z\cdots z$}}_{j}}(x_{j},y_{j},z,w)$ 
representing the model in this patch 
is given by
\beqa
\Phi_{\underbrace{\mbox{\scriptsize $z\cdots z$}}_{j}}(x_{j},y_{j},z,w)
&=&
-y_j^2+x_j^3 z^j 
+{\textstyle \frac14} (b_{2,0}+b_{2,1}z+\cdots) x_j^2 \n
&&
+ {\textstyle \frac12}  (b_{4,k}z^{k-j}+b_{4,k+1}z^{k-j+1}+\cdots)x_j 
\n
&&
+ {\textstyle \frac14} (b_{6,2k}z^{2(k-j)}+b_{6,2k+1}z^{2(k-j)+1}+\cdots)
\label{PhizjIn}
\\
&\stackrel{z\rightarrow 0}{\rightarrow}
&-y_j^2+{\textstyle \frac14} b_{2,0} x_j^2
\nonumber
\eeqa
where the exceptional ``curve'' (in the $\PP^2$ blown up over some point 
of the base with fixed (generic) $w$) splits into two lines in the sense of Kodaira.
Thus, for each generic $w$, 
$p_j$ is located at the intersection
point of these exceptional curves that have arisen from
blowing up $p_{j-1}$ $(j=1,\ldots,k-1)$.
Blowing up the final singularity $p_{k-1}$ 
yields a single irreducible exceptional curve for the $I_{2k}$ case, and 
a pair of split lines for the $I_{2k+1}$ case (see figures
\ref{fig:I2kblowup},\ref{fig:splitandover-splitI2k+1}
). 
Putting them all together, 
they constitute the $A_{2k-1}$ 
and $A_{2k}$ Dynkin diagrams 
as their intersection diagrams, as is well known.

\subsection{Conifold singularities associated with 
the split/non-split transition in the $I_{2k}$ models}
\label{subsection:ConifoldsingularitiesI2k}
Now let us explain what ``conifold singularities associated 
with the split/non-split transition'' are, by taking 
$I_{2k}$ models as an example.
Since there is no distinction between split and non-split 
fiber types in the fiber type $I_2$, let us consider $I_{2k}$ for 
$k\geq 2$.

The equation of the {\em split} $I_{2k}$ model for $k\geq 2$ 
is given by the equation (\ref{I_2kI_2k+1}) with 
\beqa
b_{2,0}=c_{1,0}^2
\label{b20=c10^2}
\eeqa
for some section $c_{1,0}$. 
A split $I_{2k}$ model exhibits, 
in addition to these ``codimension-one'' singularities,  
conifold singularities on singular 
fibers over some special loci on the base of the elliptic 
fibration, where the generic $A_{n-1}$ singularity 
is enhanced to some higher-rank one.

The discriminant of  (\ref{I_2kI_2k+1}) with (\ref{b20=c10^2}) 
reads 
\beqa
\Delta&=&\frac1{16} c_{1,0}^4 b_{8,2k}z^{2k}+\cdots.
\label{DeltaI2k}
\eeqa
$f$ and $g$ (\ref{fandg}) derived from 
(\ref{I_2kI_2k+1}) are
\beqa
f&=&-\frac1{48}c_{1,0}^4+\cdots,\n
g&=&\frac1{864}c_{1,0}^6+\cdots.
\label{fandgI2k}
\eeqa
(\ref{DeltaI2k}) shows that at the zero loci of $c_{1,0}$ 
and $b_{8,2k}$, the singularity is enhanced from $A_{2k-1}$.
Since (\ref{fandgI2k}) implies that the vanishing orders of 
$f$ and $g$ are unchanged at the zero loci of $b_{8,2k}$, 
they are ``$A_{2k}$ points'', 
which means that they are the places on the base over which 
the singularities of the fibers are enhanced to $A_{2k}$.
On the other hand, at the zero loci of $c_{1,0}$, it turns 
out that the vanishing orders of $f$ , $g$ and $\Delta$ 
go up to two, three and 
$2k+2$, 
so the zero loci of $c_{1,0}$ are  ``$D_{2k}$ points'',
 which similarly means that the singularities are enhanced to 
 $D_{2k}$ there. In fact, they are singularities of the type of 
the ``complete resolution'' \cite{MT}, meaning that they develop 
the necessary amount of conifold singularities to yield the 
degrees of freedom of matter hypermultiplets arising there.
Thus, according to the general rule \cite{KatzVafa},
the zero loci of $b_{8,2k}$ are the places (on the base) 
where a hypermultiplet transforming in ${\bf 2k}$ 
of $A_{2k-1}$ arises, and those of $c_{1,0}$ 
are where a hypermultiplet in  ${\bf k(2k-1)}$ appears. 
In general, a section $c_{i,j}$ or $b_{i,j}$ or whatever with a 
subscript $(i,j)$ is expressed as a polynomial of degree $(i-j)n+2i$ 
in $w$ \cite{MizoguchiTaniLooijenga}, so we have $(8-2k)n+16$ 
hypermultiplets in the ${\bf 2k}$ representation, 
and $n+2$ hypermultiplets in the ${\bf k(2k-1)}$ 
representation.

We  will focus on the singularity enhancement to $D_{2k}$ 
at the zero loci of $c_{1,0}$ 
since it is this singularity enhancement 
that its associated conifold singularities and their transitions
are closely related to the split/non-split transitions in F-theory.
Indeed, if we do {\em not} impose the condition (\ref{b20=c10^2}) 
to (\ref{I_2kI_2k+1}), we have an equation of the {\em non-split} 
$I_{2k}$ model, for which the corresponding $f$, $g$ and $\Delta$ 
are the ones obtained by simply replacing every $c_{1,0}^2$ with 
$b_{2,0}$ in (\ref{fandgI2k}) and (\ref{DeltaI2k}).
Even then, the vanishing orders of $f$, $g$ and $\Delta$ 
at the zero loci of $b_{2,0}$ 
remain the same as those at the loci of $c_{1,0}$,
which means that the number of $D_{2k}$ points is doubled 
($b_{2,0}$ is represented as a polynomial of degree $2n+4$ 
in $w$).

Of course, in this process of the transition from the split 
model to the non-split one, the $D_{2k}$ points, which have 
doubled in number, cannot continue to produce ${\bf k(2k-1)}$'s after 
the transition to the non-split side;
they are too many to satisfy the anomaly cancellation condition.
Therefore, the structure of the conifold singularities that existed 
before the transition to the non-split model must change after the transition.
They are what we call the conifold singularities associated with 
the split/non-split transition.
In contrast, singularity structures of the fibers over the 
$A_{2k}$ points at which $b_{8,2k}$ vanishes 
do not change by the replacement $c_{1,0}^2\leftrightarrow b_{2,0}$.\footnote{
 The six-dimensional F-theory models with an unbroken $A_5$ or 
 $A_7$ gauge symmetry also allow $E_6$ or $E_8$ points, 
but it is known \cite{KMSNS,AGRT} that they cannot be 
realized in Tate's or Deligne forms with maximal 
Tate's orders, but require to be formulated 
in a Weierstrass form or Tate's form with lower Tate's orders.
In any case, however, these singularities also do not change 
by the replacement $c_{1,0}^2\leftrightarrow b_{2,0}$ and hence have nothing to do with
 the split/non-split transition.
}

\subsection{Conifold singularities in the split $I_{2k}$ models for $k\geq 3$}
To show how these conifold singularities arise 
at the $D_{2k}$ points 
in the blowing-up process of the split $I_{2k}$ models, 
let us consider the $j$-times blown-up equation 
$\Phi_{\underbrace{\mbox{\scriptsize $z\cdots z$}}_{j-1}x}(x_{j-1},y_{j},z_{j},w)=0$
in the patch 
${\mbox{\boldmath $j$}}_{\underbrace{\mbox{\scriptsize $z\cdots z$}}_{j-1}x}$ 
 for $j=2,\ldots,k-1$ with $k\geq 3$
which is recursively defined 
in (\ref{Injthblowupzx}) in section \ref{subsection:generalities}. 
$k=2$ ($I_4$) is a special case, so we will consider it separately 
in the next subsection.

The left-hand side of this equation is explicitly given by
\beqa
\Phi_{\underbrace{\mbox{\scriptsize $z\cdots z$}}_{j-1}x}(x_{j-1}, y_j, z_j,w)
&=&
-y_j^2+x_{j-1}^j z_j^{j-1} \n
&&
+{\textstyle \frac14}  
(c_{1,0}^2+b_{2,1} x_{j-1} z_j +\cdots) \n
&&
+ {\textstyle \frac12} 
x_{j-1}^{k-j} z_j^{k-j+1}
 (b_{4,k}+b_{4,k+1}x_{j-1} z_j+\cdots) 
\n
&&
+ {\textstyle \frac14} 
x_{j-1}^{2(k-j)} z_j^{2(k-j+1)}
(b_{6,2k}+b_{6,2k+1} x_{j-1} z_j+\cdots)
\n
&=&
-y_j^2+{\textstyle \frac14}c_{1,0}^2
+x_{j-1} z_j
\left(\rule{0ex}{2.5ex}
x_{j-1}^{j-1} z_j^{j-2} 
+{\textstyle \frac14}  
b_{2,1} 
\right.
\n
&&\left.
+ {\textstyle \frac12} b_{4,k}
x_{j-1}^{k-j-1} z_j^{k-j}
+ {\textstyle \frac14} b_{6,2k}
x_{j-1}^{2(k-j)-1} z_j^{2(k-j)+1}
+O(x_{j-1} z_j)\right).\n
\label{PhizzzxI2k}
\eeqa
In general, a conifold is defined in $\CC^4 \ni (z_1,z_2,z_3,z_4)$ 
by the equation
\beqa
z_1 z_4+ z_2 z_3 =0, 
\label{conifoldeq}
\eeqa
where $(z_1,z_2,z_3,z_4)=(0,0,0,0)$ is the conifold singularity. 
Thus (\ref{PhizzzxI2k}) shows that the geometry near 
$y_j=c_{1,0}=x_{j-1}=z_j=0$ is locally approximated by 
that of a conifold, and the point itself is the conifold singularity 
for each $j=2,\ldots,k-1$ ($k\geq 3$).

Since these $k-2$ conifold singularities arise in the blowing-up process
of a split $I_{2k}$ model at each zero locus of $c_{1,0}$, 
the number of which is $n+2$ in total in the present $\FF_n$ case
(because $c_{1,0}$ is a polynomial of degree $n+2$; 
see subsection \ref{subsection:ConifoldsingularitiesI2k}.). 
Let us pay attention to a particular zero of this $c_{1,0}$, and 
we can take it to $w=0$ without loss of generality. That is, 
\beqa
c_{1,0}=w+O(w^2)
\eeqa
near $w=0$.
Then we see from  (\ref{PhizzzxI2k}) that the 
equation $\Phi_{\underbrace{\mbox{\scriptsize $z\cdots z$}}_{j-1}x}(x_{j-1}, y_j, z_j,w)=0$ 
near $(x_{j-1}, y_j, z_j,w)=(0,0,0,0)$ is
\beqa
-y_j^2+{\textstyle \frac14}w^2
+\mbox{(const.$\times$)}x_{j-1} z_j&=&0
\label{localconifoldeq}
\eeqa
up to higher-order terms.
The first two terms are factorized to yield the standard conifold equation (\ref{conifoldeq}).

The equation (\ref{localconifoldeq}) tells us that it is precisely the fact that 
the section $b_{2,0}$  is in the form of a square $c_{1,0}^2$ that the blown-up equations 
$\Phi_{\underbrace{\mbox{\scriptsize $z\cdots z$}}_{j-1}x}(x_{j-1}, y_j, z_j,w)=0$ 
give rise to conifold singularities.
If $b_{2,0}$ were not in square form $c_{1,0}^2$, which implies that the 
model is non-split,  (\ref{PhizzzxI2k}) would be
\beqa
\Phi_{\underbrace{\mbox{\scriptsize $z\cdots z$}}_{j-1}x}(x_{j-1}, y_j, z_j,w)
&=&
-y_j^2+{\textstyle \frac14}b_{2,0}
+x_{j-1} z_j
\left(\cdots
\right),
\label{nonsplitPhizzzxI2k}
\eeqa
in which $b_{2,0}$ generically vanishes like $w$ near $w=0$,
and the corresponding local equation would be
\beqa
-y_j^2+{\textstyle \frac14}w
+\mbox{(const.$\times$)}x_{j-1} z_j&=&0
\label{nonsplitlocaleq}
\eeqa
up to higher-order terms, which is not a conifold equation.
\\
\begin{figure}[h]
   \centering
   \includegraphics[width=400pt]{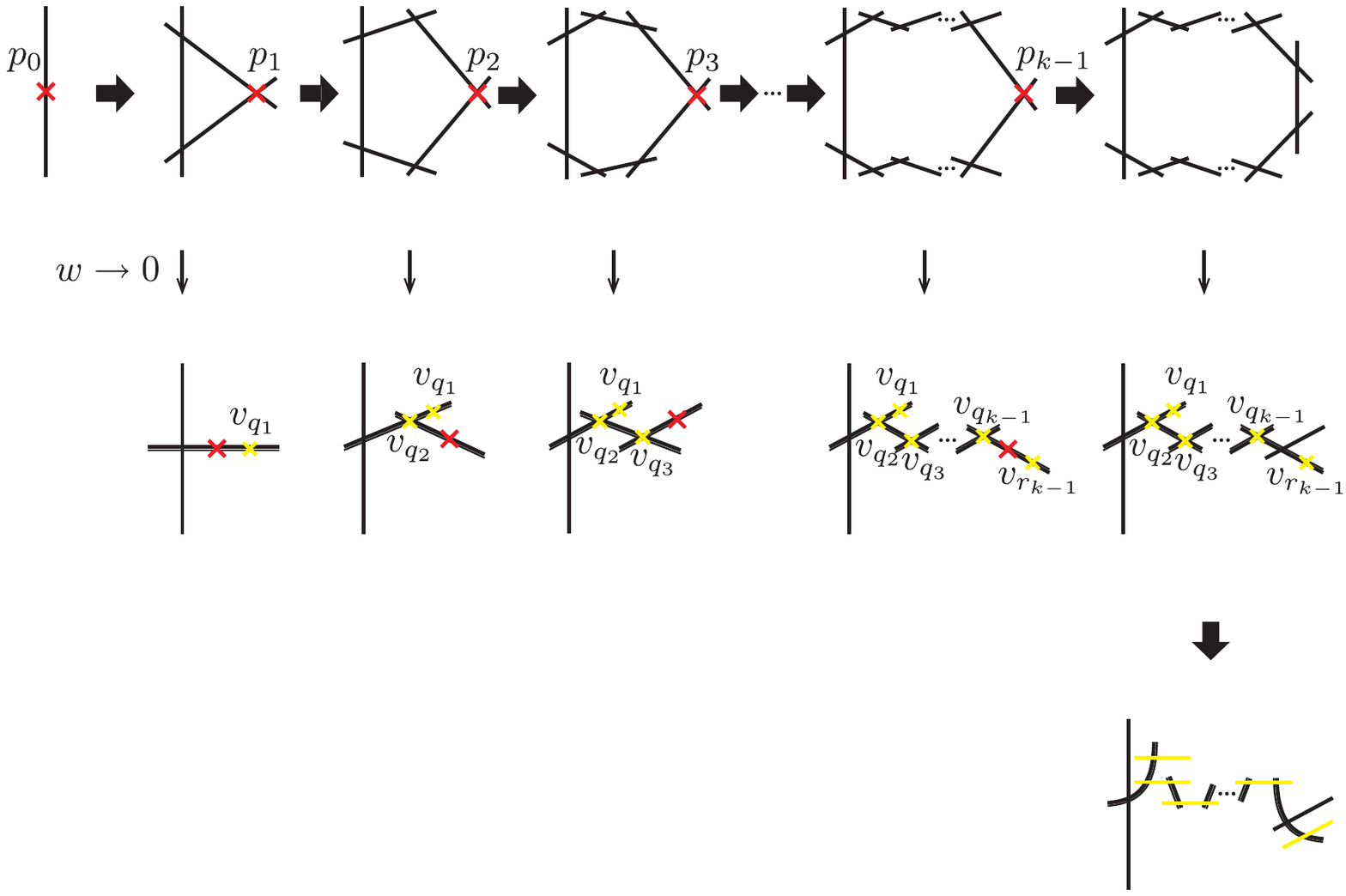} 
  \caption{Singularities and exceptional curves arising 
  in the blow-up of a split $I_{2k}$ model near a $D_{2k}$ point $w=0$. 
 ``Codimension-one'' singularities and conifold 
  singularities    
  are depicted with red and yellow x's, respectively. 
  Each bold horizontal arrow indicates a blow-up at a ``codimension-one'' 
  singularity, and the final thick downward arrow means small 
  resolutions of all the conifold singularities. The thin downward arrows 
  denote the $w\rightarrow 0$ limit.
   The left-most vertical line in each figure represents the original 
  singular fiber. }
   \label{fig:I2kblowup}
\end{figure}

In the following, we will refer to the $k-2$ conifold singularities 
arising at each zero locus of $c_{1,0}$ as\footnote{At first glance, 
this way of naming the conifold singularities 
may seem strange, but as we will see later, 
its subscript $q_j$ 
denotes the corresponding ``codimension-one''  $D_{2k}$ singularity. 
We will use ``$v$'' to denote that it is a conifold singularity.}
\beqa
v_{q_2}:&&(x_1,y_2,z_2,w)=(0,0,0,0)~~~({\mbox{\boldmath $2$}}_{zx}),\n
&\vdots&\n
v_{q_j}:&&(x_{j-1},y_j,z_j,w)=(0,0,0,0)~~~({\mbox{\boldmath $j$}}_{\underbrace{\mbox{\scriptsize $z\cdots z$}}_{j-1}x}),\n
&\vdots&\n
v_{q_{k-1}}:&&(x_{k-2},y_{k-1},z_{k-1},w)=(0,0,0,0)~~~({\mbox{\boldmath $(\!k-1\!)$}}_{\underbrace{\mbox{\scriptsize $z\cdots z$}}_{k-2}x}).
\eeqa
They are depicted with a yellow x in figure  \ref{fig:I2kblowup}.

In addition to the $k-2$ conifold singularities $v_{q_2},\ldots,v_{q_{k-1}}$, 
there are two more conifold singularities. 
One is the one on the locus of the one-time blown-up equation
$\Phi_z(x_1, y_1, z ,w)=0$ given by (\ref{PhizjIn}) with $j=1$, where  
$b_{2,0}$ satisfies the split condition $b_{2,0}=c_{1,0}^2$.
If $k\geq 3$, 
$\Phi_z(x_1, y_1, z ,w)$ can be written as
\beqa
\Phi_z(x_1, y_1, z ,w)
&=&
-y_1^2+{\textstyle \frac14}c_{1,0}^2 x_1^2
+z
\left(\rule{0ex}{2.5ex}
x_1^3  
+{\textstyle \frac14}  
b_{2,1} x_1^2+O(z)
\rule{0ex}{2.5ex}\right),
\label{PhizI2k}
\eeqa
so focusing on a particular zero of $c_{1,0}$ and set  
$c_{1,0}=w$, 
the equation becomes 
\beqa
-y_1^2+{\textstyle \frac14}w^2 x_1^2
+z
\left(\rule{0ex}{2.5ex}
x_1^3  
+{\textstyle \frac14}  
b_{2,1} x_1^2
\rule{0ex}{2.5ex}\right)&=&0
\eeqa
near  $z=w=0$.
$y_1=w=z=x_1=0$ is a special case of $ p_1$, so 
assuming $x_1\neq 0$, we find 
 \beqa
 v_{q_1}:&&(x_1,y_1,z,w)=(-{\textstyle \frac14}  
b_{2,1},0,0,0)~~~({\mbox{\boldmath $1$}}_{z})
\eeqa
is a conifold singularity that arises besides $v_{q_2},\ldots,v_{q_{k-1}}$.

The other conifold singularity can be found on the locus of 
$\Phi_{\underbrace{\mbox{\scriptsize $z\cdots z$}}_{k-1}}(x_{k-1},y_{k-1},z,w)$,
which is given by (\ref{PhizjIn}) with setting $j=k-1$. 
We  have already discussed that it  
has a codimension-one singularity $p_{k-1}$ at $(x_{k-1},y_{k-1},z,w)=(0,0,0,w)$.
We can show that it also has a conifold singularity if $b_{2,0}=c_{1,0}^2$ 
for some $c_{1,0}$ by writing, for $k\geq 3$,
\beqa
\Phi_{\underbrace{\mbox{\scriptsize $z\cdots z$}}_{k-1}}(x_{k-1},y_{k-1},z,w)
&=&
-y_{k-1}^2
+{\textstyle \frac14} c_{1,0}^2 x_{k-1}^2\n
&&
+z\left(
x_{k-1}^3 z^{k-2} 
+
{\textstyle \frac14} b_{2,1} x_{k-1}^2 
+ {\textstyle \frac12} b_{4,k} x_{k-1}
+O(z)
\right).
\eeqa
Thus, by setting $c_{1,0}=w$, the blown-up equation is reduced 
near  $z=0$ to
\beqa
-y_{k-1}^2+{\textstyle \frac14}w^2 x_{k-1}^2
+z
\left(\rule{0ex}{2.5ex}
{\textstyle \frac14} b_{2,1} x_{k-1}^2 
+ {\textstyle \frac12} b_{4,k} x_{k-1}
\rule{0ex}{2.5ex}\right)&=&0,
\eeqa
which shows that 
 \beqa
 v_{r_{k-1}}:&&(x_{k-1},y_{k-1},z,w)=
 \left(-\frac{2b_{4,k}}{b_{2,1}},0,0,0\right)~~
 ({\mbox{\boldmath $(k-1)$}}_{\underbrace{\mbox{\scriptsize $z\cdots z$}}_{k-1}})
 \label{vrk-1}
\eeqa
is another conifold singularity. 

Thus, the split $I_{2k}$ model gives rise to a total of $k-2+2=k$ 
conifold singularities at each zero locus of $c_{1,0}$.
They are resolved by small resolutions to give $k$ exceptional curves, 
and comprise, 
together with the $k$ exceptional curves coming from the codimension-one  
singularities, the $D_{2k}$ Dynkin diagram (figure \ref{fig:I2kblowup}).

\subsection{Conifold singularities in the split $I_{4}$ model (the $k=2$ case)}
Although similar, the split $I_4$ model, 
which is the lowest $k(=2)$ case,
is slightly different from the models for $k\geq 3$ 
in the way the conifold singularities appear, so 
we will briefly comment on this special case for completeness.

We have seen that in a split $I_{2k}$ model with $k\geq 3$, 
two special conifold singularities $v_{q_1}$ and $v_{r_{{k-1}}}$ 
appear in the patches ${\mbox{\boldmath $1$}}_{z}$ and  
${\mbox{\boldmath $(k-1)$}}_{\underbrace{\mbox{\scriptsize $z\cdots z$}}_{k-1}}$, 
respectively. If $k=2$, they are the same patches. 
Therefore, in the $k=2$ case, there appear both conifold 
singularities on the zero locus of $\Phi_z(x_1, y_1, z ,w)$
defined in $({\mbox{\boldmath $1$}}_{z})$, in addition to 
the ``codimension-one'' singularity $p_1$. 
After the resolutions, they yield the $D_4$ Dynkin 
diagram as their intersection diagram.

\subsection{Split/non-split transitions as conifold transitions in the $I_{2k}$ 
models}
Now, we can 
discuss the relationship between the split/non-split 
transition and the conifold transition. 
To summarize what we have learned so far about the $I_{2k}$ model:
\begin{itemize}
\item{If $b_{2,0}$ is a square of some $c_{1,0}$, the model is 
split, otherwise non-split. }
\item{In the split models, $D_{2k}$ points are $n+2$ double roots 
of the $(2n+4)$th order equation $b_{2,0}=c_{1,0}^2=0$ of $w$, 
while in the non-split models, they are generically $2n+4$ single roots.}
\item{In the split case, there arise $k$ conifold singularities 
at each zero locus of $c_{1,0}$, while in the non-split case, no 
conifold singularities appear at the loci of $b_{2,0}$.}
\end{itemize}
So let us consider a deformation of the complex structure
(of the total elliptic fibration) 
in which a particular double root, say $w=0$,  ``splits'' into
two single roots $w=\pm\epsilon$ that are minutely 
separated $|\epsilon|\ll1$. By deforming just one of the  
$n+2$ double roots into a pair of single roots, $b_{2,0}$ 
can no longer be written in the form of a square of anything,
so this deformation turns the split model into a non-split model. 
This deformation is achieved by replacing $w^2$ with 
$w^2-\epsilon^2$, and turns 
the conifold
\beqa
-y^2+w^2+xz=0
\eeqa
into
\beqa
-y^2+w^2+xz=\epsilon^2,
\eeqa
which is the {\em deformed conifold}\,!

One can easily verify that all the conifold singularities 
$v_{q_1},\ldots,v_{q_{k-1}}, v_{r_{k-1}}$ are deformed 
into local deformed conifolds\footnote{By a ``local conifold'' 
we mean the geometry near the conifold singularity described 
by an equation $(z_1 z_4+ z_2 z_3)(1+O(z_i)) =0$. Similarly, 
by a ``local deformed conifold'' we mean 
the one described 
by  $(z_1 z_4+ z_2 z_3-\epsilon^2)(1+O(z_i)) =0$. 
} 
by the replacement 
$w^2\rightarrow w^2-\epsilon^2$. This means that 
the special deformation of the complex structure of the 
total elliptic fibration that makes a double zero of $w$ split 
into a pair is exactly the deformation of the complex structure 
of the local conifolds.

Suppose that we start from a singular split $I_{2k}$ model 
given by the equation (\ref{I_2kI_2k+1}), where $b_{2,0}=c_{1,0}^2$, 
and $b_{2k,8}$ does not vanish. By blowing up all the ``codimension-one'' 
singularities of it, we end up with a geometry whose only singularities 
are conifold singularities. There are two ways to smooth these singularities.
One is to resolve them by small resolutions; this just yields a smooth 
split $I_{2k}$ model. The other is to deform the conifold singularities; 
this is achieved by replacing  $b_{2,0}=c_{1,0}^2$ with 
$b_{2,0}=c_{1,0}^2-\epsilon_{1,0}^2$ for some section $\epsilon_{1,0}$, 
then the model is a smooth {\em non-split} $I_{2k}$ model. 
In other words, the split/non-split transition in an $I_{2k}$ model is 
nothing but a conifold transition.

As we have seen above, there is not just one conifold 
singularity that appears at each zero locus of $c_{1,0}$ 
and is involved in the transition. 
There are $k$ such conifold singularities at each locus, 
and they are simultaneously deformed to give a 
non-split model.

\subsection{The mechanism proposed by \cite{AKM} for non-local matter generation }
%

As mentioned in Introduction, the origin of non-local matter was proposed 
\cite{AKM} as due to the adjoint hypermultiplets associated with a certain 
genus-$g$ curve in the elliptically fibered CY3. 
In this section, let's see how their proposal can be actually 
implemented in the blowing-up process we have discussed so far.

In general, fiber degeneration occurs at a codimension-one  
discriminant locus on the base, which is a curve on the two-dimensional base 
($\FF_n$ in our case) of the CY3.
Thus, together with the degenerate $\PP^1$ fiber, 
with a possible singularity 
before blowing up, it forms a ruled (=$\PP^1$-fibered) 
surface in the CY3. 
We are interested in the gauge 
divisor, over which there is a distinction between the split or the non-split
fiber type. 

Since we take the gauge divisor to be a divisor of 
the {\em fiber} $\PP^1$ of the Hirzebruch surface $\FF_n$ (that is, $z=0$), 
we may naturally take the base of the ruled surface 
to be the {\em base} $\PP^1$ of the $\FF_n$ (parametrized by $w$),
which was called $M_1$ in \cite{AKM}.
Its genus is $0$; this agrees with
\cite{Sadov}, in which, by an anomaly analysis, 
the number of the adjoint hypers was shown  
to coincide with the genus of the gauge divisor, 
and the fact that there is no massless adjoint hypermultiplet
in the spectrum \cite{BIKMSV}.

The proposal of \cite{AKM} was as follows:
Taking a non-split $I_{2k}$ model as an example, if the singularity 
of the $\PP^1$ fiber of the ruled surface is blown up, 
the singular point at each fixed $w$ is replaced by a collection of $\PP^1$'s, 
which form (over the whole base) a smooth surface 
consisting of multiple components corresponding to 
different nodes of the $A_{2k-1}$ Dynkin diagram. 
In the non-split case these $\PP^1$'s (exceptional fibers)
are merged in pairs smoothly, except for the one corresponding to the middle node. 
This is precisely why the gauge algebra is reduced to a non-simply-laced one by the identification under the diagram automorphism, but in \cite{AKM} 
they further note that a component of 
the surface swept by a particular pair of such exceptional 
fibers is also a ruled surface, 
whose base is a 2-sheeted Riemann surface of genus $g$.
This genus-$g$ base, called $M_2$ in \cite{AKM}, is a double cover of $M_1$
and has $2g+2$ branch points over which the pair of exceptional fibers 
meet and join smoothly in the non-split model.
\cite{AKM} argued that, according to \cite{Witten,KMP}, $g$ hypermultiplets  
arise from the harmonic 1-forms of the genus-$g$ Riemann surface and are assigned 
to one of the short simple roots of the $C_{k}$ Dynkin diagram.

Let us consider how this genus-$g$ Riemann surface 
can be seen in our set-up.
We could consider the general equation for $I_{2k}$ given in section 3.4, 
but to simplify the notation and clarify the issue, we will instead repeat 
the blow-up procedure with the {\em homogeneous} coordinates in the $I_6$ model, 
 the simplest case where there are more than one  pair of exceptional curves identified by monodromy.

Again, starting from equation (\ref{I_2kI_2k+1}), let $k=3$.
This time, instead of (\ref{patches1}), we change the coordinates as
\beqa
(x,y,z)&=&(\alpha x_1, \alpha y_1, \alpha z_1),
\label{homogeneous}
\eeqa
where  $(x_1:y_1:z_1)$ are homogeneous coordinates of $\PP^2$ and 
$\alpha\in \CC$. Plugging (\ref{homogeneous}) into $\Phi(x,y,z,w)$, 
we define 
\beqa
 \alpha^{-2}\Phi(\alpha x_1, \alpha y_1, \alpha z_1,w)&\equiv&
 \Phi_\alpha(x_1,y_1,z_1,\alpha,w)\n
 &=&-y_1^2+x_1^3 \alpha +{\textstyle \frac14} (b_{2,0}+b_{2,1}z_1 \alpha+\cdots) x_1^2 \n
&&
+ {\textstyle \frac12}  (b_{4,3}z_1^3 \alpha^2+\cdots)x_1 
\n&&
+{\textstyle \frac14} (b_{6,6}z_1^{6}\alpha^4+\cdots),
\label{I6homogeneous1stblowup}
\eeqa
similarly to (\ref{In1stblowup}).
Of course, if $z_1=1$ and $\alpha$ is renamed $z$,  
$\Phi_\alpha(x_1,  y_1, z_1=1,\alpha =z,w)$
becomes $\Phi_z(x_1,  y_1, z,w)$ (\ref{PhizjIn}) 
with $j=1$, $k=3$.
As we discussed in the previous section, if the section $b_{2,0}$ is a 
deformation of a square $c_{1,0}^2$, the equation $\Phi_\alpha(x_1,y_1,z_1,\alpha,w)=0$ 
describes a three-manifold with $n+2$ deformed conifold ``singularities'' near the zero loci 
of $c_{1,0}$. The exceptional curves can be found 
at the intersection with the divisor $\alpha=0$:
\beqa
\Phi_\alpha(x_1,y_1,z_1,\alpha=0,w)&=&-y_1^2+{\textstyle \frac14} b_{2,0}(w)x_1^2 ~~=0,
\label{genusgbase1}
\eeqa
where we have recovered the argument of $b_{2,0}$ to remember that it is a polynomial of degree $2n+4$ in $w$.
With fixed $w$, (\ref{genusgbase1}) represents a pair of $\PP^1$'s in
$\PP^2\ni (x_1:y_1:z_1)$ if $b_{2,0}(w)\neq 0$
intersecting at  $(x_1:y_1:z_1)=(0:0:1)$, which is a singularity 
to be blown up in the next step, thereby it is to be separated into 
two distinct points on the respective two $\PP^1$'s.
Thus if the value of $w$ is varied, the two $\PP^1$'s as a whole 
yield a surface, which comprises $S_2$ in \cite{AKM}.

On the other hand, (\ref{genusgbase1}) can also be viewed 
as a 2-sheeted Riemann surface, and, by ``forgeting'' $z_1$, 
any point on this (component of the) surface $S_2$ 
has a unique projection onto this Riemann surface.
Therefore, it is a ruled surface whose base 
is a 2-sheeted Riemann surface given by (\ref{genusgbase1})
(provided that $(x_1:y_1:z_1)=(0:0:1)$ is blown up), 
which may be called $M_2$ in the notation of \cite{AKM}.

However, another similar Riemann surface arises 
in the next step of the blow-up. Since 
 $\Phi_\alpha(x_1,y_1,z_1,\alpha,w)=0$ 
 is singular at $(x_1:y_1:z_1)=(0:0:1)$, we blow up there 
 by defining 
 \beqa
(x_1,y_1,\alpha)&=&(\beta x_2, \beta y_2, \beta \alpha_2),
\label{homogeneous2}
\eeqa
where $(x_2:y_2:\alpha_2)$ are also 
homogeneous coordinates of $\PP^2$ and 
$\beta\in \CC$. Plugging (\ref{homogeneous2}) into 
$\Phi_\alpha(x_1,y_1,z_1,\alpha,w)$, 
we similarly obtain 
\beqa
 \beta^{-2}\Phi_\alpha(\beta x_2, \beta y_2, z_1, \beta\alpha_2,w)
 &\equiv&
 \Phi_{\alpha\beta}(x_2,y_2,z_1,\alpha_2,\beta,w)\n
 &=&-y_2^2+x_2^3 \alpha_2 \beta^2 
 +{\textstyle \frac14} (b_{2,0}+b_{2,1}z_1 \alpha_2\beta+\cdots) x_2^2 \n
&&
+ {\textstyle \frac12}  (b_{4,3}z_1^3 \alpha_2^2 \beta+\cdots)x_2 
\n&&
+{\textstyle \frac14} (b_{6,6}z_1^{6}\alpha_2^4 \beta^2+\cdots).
\label{I6homogeneous2ndblowup}
\eeqa
The exceptional curves are 
at the intersection with the divisor $\beta=0$:
\beqa
\Phi_{\alpha\beta}(x_2,y_2,z_1,\alpha_2,\beta=0,w)&=&-y_2^2+{\textstyle \frac14} b_{2,0}(w)x_2^2 ~~=0.
\label{genusgbase2}
\eeqa
This is again a ruled surface (without any further blowing up), whose base is also a Riemann surface given by the same equation 
(\ref{genusgbase2}) with $\alpha_2$ forgotten. 

Clearly, (\ref{genusgbase1}) and (\ref{genusgbase2}) are different components of the ruled 
surface $S_2$, residing on different divisors $\alpha=0$ and $\beta=0$, 
respectively. The important point, however, is that they represent the 
{\em same} Riemann surface as the base space. Indeed, 
for a given $w$, (\ref{genusgbase1}) and (\ref{genusgbase2}) 
respectively determine the ratios $x_1:y_1$ and $x_2:y_2$, 
but they are the same by definition and are consistent.
Thus we may successfully say that $S_2$ is a ruled surface 
over {\em a} genus-$g$ ($=n+1$ here) Riemann surface $M_2$,
as \cite{AKM} claimed.

It is also straightforward to check that, for general $I_{2k}$ $(k\geq 3)$ 
models defined by (\ref{I_2kI_2k+1}),
all the genus-$g$ bases that appear at each blow-up are 
the same (except at the final blow-up where such a genus-$g$ 
curve does not arise). Similar holds for the $I_{2k+1}$ 
non-split models\footnote{In this case, the exceptional curve arising at the final blow-up 
splits into two lines, but still the genus-$g$ Riemann surfaces arising before the final blow-up are all identical.}. 
In the non-split $I_n^*$ and $IV$ models, 
since there is only one pair of exceptional curves identified by 
monodromy, the problem described above does not arise.
Finally, it can be verified that the two genus-$g$ bases 
appearing in the non-split $IV^*$ model are also the same. 

Thus we have seen that, even when there are multiple pairs of exceptional curves and $S_2$ consists of multiple components, the genus-$g$ Riemann surface $M_2$ is well defined and serves the mechanism proposed by \cite{AKM}.

\section{Split/non-split transitions as conifold transitions (II): the $I_{2k+1}$ 
models}
Although the defining equations of 
the $I_{2k}$ and $I_{2k+1}$ models 
are common (\ref{I_2kI_2k+1}), the relationship between the 
split/non-split transition and the conifold transition in the $I_{2k+1}$ models 
is quite different from that in the $I_{2k}$ models. 

The most significant difference is that 
in the split $I_{2k+1}$ model, 
the singularity (in the sense of the Kodaira fiber) is 
enhanced from $A_{2k}$ to $D_{2k+1}$ 
at the zero loci of $b_{2,0}$ (which is in the form of a square 
$c_{1,0}^2$ for some $c_{1,0}$), 
whereas in the non-split model, the 
singularity at the generic zero loci of $b_{2,0}$ is enhanced to 
$D_{2k+2}$
instead of to $D_{2k+1}$. Consequently, a generic split $I_{2k+1}$ model 
does not directly transition to a non-split $I_{2k+1}$ model. Rather, we 
will show that there is a certain special interface model that connects 
the split and non-split $I_{2k+1}$ models via a conifold transition. 

\begin{figure}[ht]
   \centering
   \includegraphics[width=270pt]{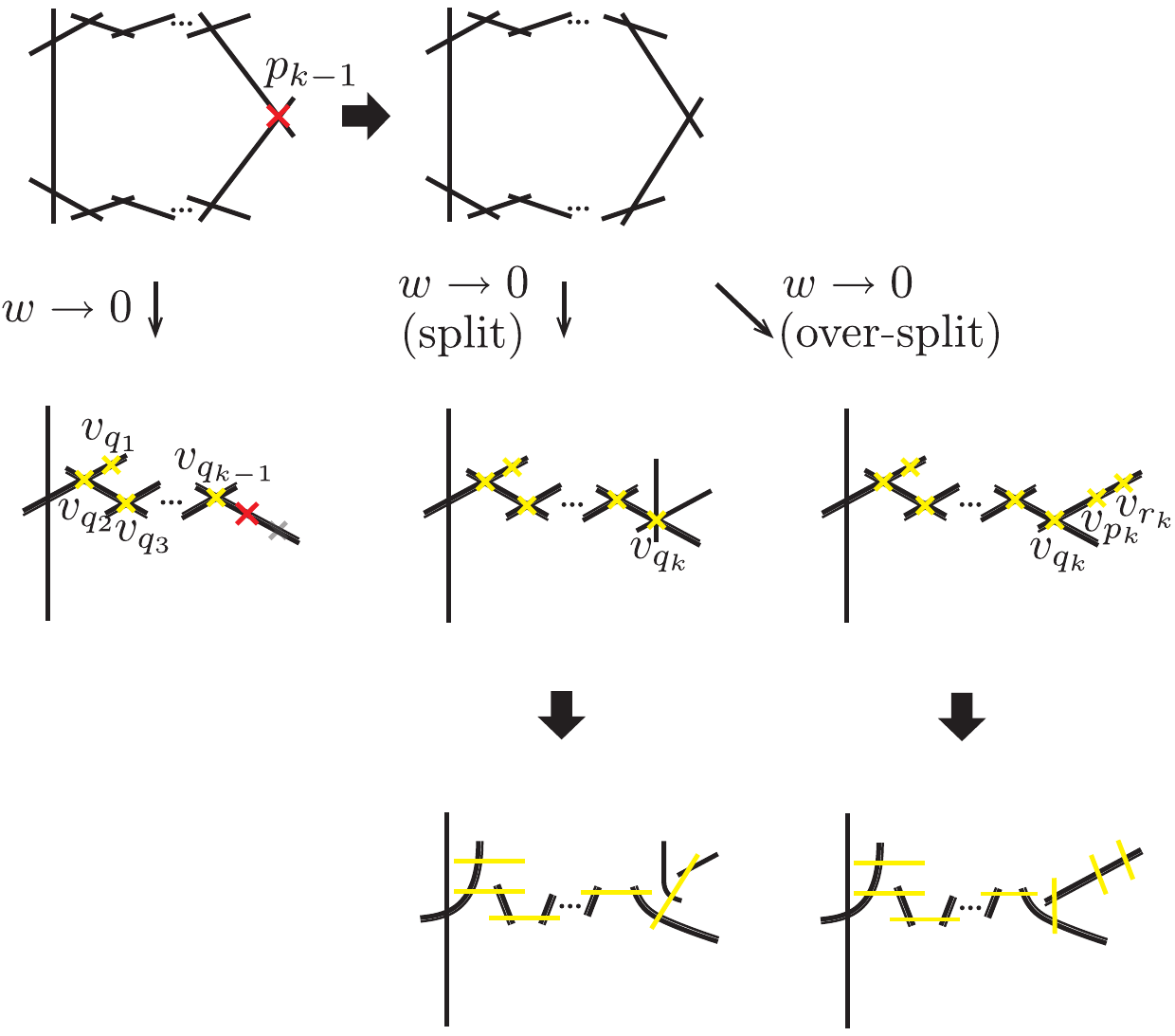} 
  \caption{Singularities and exceptional curves 
  in a split and an over-split $I_{2k+1}$ model for $k\geq 2$ 
  near a double root of $c_{1,0}^2=0$. }
   \label{fig:splitandover-splitI2k+1}
\end{figure}

\subsection{The split, non-split and ``over-split'' $I_{2k+1}$ models}
The vanishing orders of the sections $b_2$, $b_4$, $b_6$ for a $I_{2k+1}$ 
model are $0$, $k$, $2k$, respectively, which are the same as those for a $I_{2k}$ model.
The difference from the $I_{2k}$ model is that the vanishing order 
of $b_8$ is $2k+1$ instead of $2k$, which means that
\beqa
0=4b_{8,2k}=b_{4,k}^2-b_{2,0}b_{6,2k}.
\label{b82k=0}
\eeqa

In the split models, $b_{2,0}$ is given by a square 
$c_{1,0}^2$ for some $c_{1,0}$, so we have
\beqa
b_{6,2k}=\left(\frac{b_{4,k}}{c_{1,0}}\right)^2.
\eeqa
Thus $b_{4,k}$
must be divisible by 
$c_{1,0}$.
We can then write
\beqa
b_{2,0}&=&c_{1,0}^2,\n
b_{4,k}&=&c_{1,0}c_{3,k},\n
b_{6,2k}&=&c_{3,k}^2
\label{I2k+1splitb2b4b6}
\eeqa
for some $c_{3,k}$, which is a section of 
the line bundle specified by its subscripts. 
Again, $k=1$ is a special case so 
will be discussed later. For $k\geq 2$,  
we find  
\beqa
f&=&-\frac1{48}c_{1,0}^4+\cdots\n
&\stackrel{c_{1,0}\rightarrow 0}{\rightarrow}&
-\frac1{48}b_{2,1}^2 z^2+\cdots,
\n
g&=&\frac1{864}c_{1,0}^6+\cdots\n
&\stackrel{c_{1,0}\rightarrow 0}{\rightarrow}&
\frac1{864}b_{2,1}^3 z^3+\cdots 
\label{fandgsplitI2k+1}
\eeqa
and 
\beqa
\Delta&=&\frac1{16} c_{1,0}^4 b_{8,2k+1}z^{2k+1}+\cdots
\n
&\stackrel{c_{1,0}\rightarrow 0}{\rightarrow}&
\frac1{64}b_{2,1}^3 c_{3,k}^2 z^{2k+3}+\cdots.
\label{DeltasplitI2k+1}
\eeqa
Therefore,  the zero loci of $c_{1,0}$ are where the 
apparent 
fiber type changes to $I_{2k-3}^*$, or from $A_{2k}$ to $D_{2k+1}$ 
in terms of the singularity. 
\footnote{Again, as we noted in section \ref{subsection:ConifoldsingularitiesI2k}, 
an enhancement to $E_7$  is possible in 
the F-theory model with an unbroken $A_6$ gauge symmetry, 
but it also cannot be realized in our Deligne form 
\cite{AGRT,MizoguchiTaniNonCartan}. 
It is also irrelevant for the split/non-split transition.
} 

In the non-split $I_{2k+1}$ models, (\ref{b82k=0}) is 
assumed to be satisfied, but $b_{2,0}$ is not assumed 
to be in the form of a square.
So suppose that $b_{2,0}$ is not a complete square but 
takes the product form 
\beqa
b_{2,0}&=&c_{r,0}^2\tilde b_{2-2r,0}
\eeqa
for some $c_{r,0}$ and $\tilde b_{2-2r,0}$.
In this case, $b_{4,k}$ must be divisible by $c_{r,0}$. 
Then the same discussion as we did in the split $I_{2k+1}$ 
model can apply to show that at the zero loci of $c_{r,0}$ 
the fiber type changes there to $I_{2k-3}^*$ and the 
singularity is enhanced to $D_{2k+1}$. 

Thus let us assume that $b_{2,0}$ is completely generic 
and has no square factor, that is, the equation $b_{2,0}=0$ 
has no double root. In this case, the constraint (\ref{b82k=0})
requires that $b_{4,k}$ is divisible by $b_{2,0}$:
\beqa
b_{2,0}&:&\mbox{generic,}\n
b_{4,k}&=&b_{2,0}c_{2,k},\n
b_{6,2k}&=&b_{2,0}c_{2,k}^2
\label{I2k+1nonsplitb2b4b6}
\eeqa
for some section $c_{2,k}$ of 
the line bundle implied by the subscripts. 
For $k\geq 2$, we can see that the $z$-expansions of 
$f$ and $g$ are similar to (\ref{fandgsplitI2k+1}), 
but the discriminant in the present case is
\beqa
\Delta&=&\frac1{16} b_{2,0}^2 b_{8,2k+1}z^{2k+1}+\cdots
\n
&\stackrel{b_{2,0}\rightarrow 0}{\rightarrow}&
\frac1{64}b_{2,1}^2 (b_{2,1}b_{6,2k+1}-b_{4,k+1}^2) 
z^{2k+4}+\cdots,
\label{DeltanonsplitI2k+1}
\eeqa
in which the order of $z$ at the zero loci of $b_{2,0}$ 
is one order higher than that in the split case. This shows 
that, in a non-split $I_{2k+1}$ model, the fiber type 
in the sense of Kodaira changes to  $I_{2k-2}^*$ 
instead of $I_{2k-3}^*$, and the apparent singularity 
there is enhanced from $A_{2k}$ to $D_{2k+2}$ 
instead of $D_{2k+1}$.

Therefore, a generic split $I_{2k+1}$ model cannot directly 
transition to a non-split $I_{2k+1}$ model. The interface model 
that connects the split and non-split models can be 
obtained by tuning the complex structure of a split model 
so that it can yield the $D_{2k+2}$ points which are originally 
absent in generic split $I_{2k+1}$ models. 
The existence of such models was already pointed out 
in \cite{Tani}. 
More 
specifically, we consider a special class of 
split $I_{2k+1}$ models in which the relevant sections 
$b_{2,0}$, $b_{4,k}$ and $b_{6,2k}$ are given by 
\beqa
b_{2,0}&=&c_{1,0}^2\n
b_{4,k}&=&c_{1,0}^2 c_{2,k},\n
b_{6,2k}&=&c_{1,0}^2 c_{2,k}^2,
\label{I2k+1oversplitb2b4b6}
\eeqa
which we call an ``over-split $I_{2k+1}$ model.''
(\ref{I2k+1oversplitb2b4b6}) can be obtained by
specializing $c_{3,k}$ to the factorized form $c_{1,0}c_{2,k}$ 
for some $c_{2,k}$. This in particular implies that 
$c_{3,k}$ in (\ref{DeltasplitI2k+1}) vanishes as 
$c_{1,0}\rightarrow 0$. The next non-vanishing order 
is $2k+4$, yielding the desired enhancement to $D_{2k+2}$. 
It is also clear that replacing $c_{1,0}^2$ with $b_{2,0}$ 
in (\ref{I2k+1oversplitb2b4b6}) yields the specifications 
of the sections in the non-split models (\ref{I2k+1nonsplitb2b4b6}).

\subsection{Conifold singularities in the $I_{2k+1}$ models for $k\geq 2$}
We will now blow up the ``codimension-one'' singularities of 
the split and over-split $I_{2k+1}$ models. Since the only difference 
between the $I_{2k}$ and the $I_{2k+1}$ models (in their definitions) 
is the vanishing order of $b_8$, the way the singularities are 
blown up is very similar between the two. When we blow up the ``codimension-one'' singularities of a split $I_{2k+1}$ model,  the first difference from 
the $I_{2k}$ models 
we encounter is the absence of the conifold singularity $v_{r_{k-1}}$ 
in the coordinate patch ${\mbox{\boldmath $(k-1)$}}_{\underbrace{\mbox{\scriptsize $z\cdots z$}}_{k-1}}$ 
 (\ref{vrk-1}), which appeared in the $I_{2k}$ models when $w\equiv c_{1,0}\rightarrow 0$. Instead, if we blow up the ``codimension-one'' 
 singularity $p_{k-1}$, we get a pair of exceptional curves, 
 at the intersection of which there is a conifold singularity $v_{q_k}$ 
 (figure \ref{fig:splitandover-splitI2k+1}). If we resolve all the conifold 
 singularities 
by small resolutions, 
 we obtain the $D_{2k+1}$ Dynkin diagram as the intersection 
 diagram of the resulting exceptional curves.

On the other hand, if we blow up the singularity 
$p_{k-1}$ in the over-split $I_{2k+1}$ model, the 
pair of exceptional lines come on top of each other 
to form a single irreducible line, on which 
three conifold singularities $v_{p_k},v_{q_k}$ and 
$v_{r_k}$ appear. Resolving all 
the conifold singularities gives the $D_{2k+2}$ Dynkin diagram
in this case.

How these conifold singularities arise in the blowing-up 
process of the split and over-split $I_{2k+1}$ models 
near a double root of $c_{1,0}^2=0$ 
is summarized in figure \ref{fig:splitandover-splitI2k+1}.

\subsection{The split/non-split transitions and conifold transitions 
in the $I_{2k+1}$ models for $k\geq 2$}
Again, let us focus on a particular double root of $c_{1,0}^2=0$,  
and let it be $w=0$.
Then the local equations yielding the conifold singularities 
$v_{q_1},\ldots,v_{q_{k-1}}$ are the same as those in 
the split $I_{2k}$ models. To see how the conifold singularities 
$v_{p_k},v_{q_k},v_{r_k}$ arise, let us consider the $k$-times blown-up 
equation $\Phi_{\underbrace{\mbox{\scriptsize $z\cdots z$}}_{k-1}x}(x_{k-1}, y_k, z_k,w)
=0$ in the patch 
${\mbox{\boldmath $k$}}_{\underbrace{\mbox{\scriptsize $z\cdots z$}}_{k-1}x}$,
where
\beqa
\Phi_{\underbrace{\mbox{\scriptsize $z\cdots z$}}_{k-1}x}(x_{k-1}, y_k, z_k,w)
&\equiv&
x_{k-1}^{-2}
\Phi_{\underbrace{\mbox{\scriptsize $z\cdots z$}}_{k-1}}(x_{k-1}, x_{k-1}y_k, x_{k-1}z_k,w)
\n
&=&-y_k^2 + x_{k-1}^k z_k^{k-1} \n
&&
+{\textstyle \frac14}  
(c_{1,0}^2+b_{2,1} x_{k-1} z_k +\cdots) \n
&&
+ {\textstyle \frac12} 
 (c_{1,0}c_{3,k}z_k+b_{4,k+1}x_{k-1} z_k^2+\cdots) 
\n
&&
+ {\textstyle \frac14} 
(c_{3,k}^2 z_k^2 +b_{6,2k+1} x_{k-1} z_k^3+\cdots)
\n
&\stackrel{x_{k-1}\rightarrow 0}{\rightarrow}&
-y_k^2 
+{\textstyle \frac14}  
(c_{1,0}+ c_{3,k}z_k)^2
\label{I2k+1Phizk-1x}
\eeqa
in the split case. The last line shows that the exceptional curve
splits into two lines, which intersect at 
\beqa
x_{k-1}=y_k=c_{1,0}+c_{3,k}z_k=0.
\eeqa
If $c_{1,0}=0$, $z_k$ also vanishes for generic $c_{3,k}$; this is a conifold 
singularity. Indeed, we can write 
$\Phi_{\underbrace{\mbox{\scriptsize $z\cdots z$}}_{k-1}x}(x_{k-1}, y_k, z_k,w)$
as, setting $c_{1,0}=w$,
\beqa
\Phi_{\underbrace{\mbox{\scriptsize $z\cdots z$}}_{k-1}x}(x_{k-1}, y_k, z_k,w)
&=&
-y_k^2 
+{\textstyle \frac14}  
(w+ c_{3,k}z_k)^2
+x_{k-1}z_k\left(x_{k-1}^{k-1}z_k^{k-2}\right.\n
&&\left.
+{\textstyle \frac14}b_{2,1}
+{\textstyle \frac12}b_{4,k+1}z_k
+{\textstyle \frac14}b_{6,2k+1}z_k^2
+O(x_{k-1}z_k)
\right).  
\label{I2k+1splitconifoldeq}
\eeqa
This shows that 
\beqa
 v_{q_k}:&&(x_{k-1},y_k,z_k,w)=(0,0,0,0)~~~({\mbox{\boldmath $k$}}_{\underbrace{\mbox{\scriptsize $z\cdots z$}}_{k-1}x})
\eeqa
is a conifold singularity. This is the only conifold singularity 
in this patch in the split case. Note that the $w$-dependence of 
(\ref{I2k+1splitconifoldeq}) 
is not only through $w^2$.

In the over-split case, (\ref{I2k+1Phizk-1x}) becomes
\beqa
\Phi_{\underbrace{\mbox{\scriptsize $z\cdots z$}}_{k-1}x}(x_{k-1}, y_k, z_k,w)
&=&-y_k^2 + x_{k-1}^k z_k^{k-1} \n
&&
+{\textstyle \frac14}  
(c_{1,0}^2+b_{2,1} x_{k-1} z_k +\cdots) \n
&&
+ {\textstyle \frac12} 
 (c_{1,0}^2c_{2,k}z_k+b_{4,k+1}x_{k-1} z_k^2+\cdots) 
\n
&&
+ {\textstyle \frac14} 
(c_{1,0}^2c_{2,k}^2 z_k^2 +b_{6,2k+1} x_{k-1} z_k^3+\cdots)
\n
&\stackrel{x_{k-1}\rightarrow 0}{\rightarrow}&
-y_k^2 
+{\textstyle \frac14}  
c_{1,0}^2(1+ c_{2,k}z_k)^2.
\label{I2k+1Phizk-1xoversplit}
\eeqa
Thus, the exceptional curves that are split into two lines 
at $c_{1,0}\neq 0$ overlap into a single line at  $c_{1,0}=0$.
In this case, by setting $c_{1,0}\equiv w$, 
(\ref{I2k+1Phizk-1xoversplit}) can be written as
\beqa
\Phi_{\underbrace{\mbox{\scriptsize $z\cdots z$}}_{k-1}x}(x_{k-1}, y_k, z_k,w)
&=&
-y_k^2 
+{\textstyle \frac14}  
w^2(1+ c_{2,k}z_k)^2
+x_{k-1}z_k\left(x_{k-1}^{k-1}z_k^{k-2}\right.\n
&&\left.
+{\textstyle \frac14}b_{2,1}
+{\textstyle \frac12}b_{4,k+1}z_k
+{\textstyle \frac14}b_{6,2k+1}z_k^2
+O(x_{k-1}z_k)
\right),
\label{I2k+1oversplitconifoldeq}
\eeqa
which shows that there are three conifold singularities at 
$x_{k-1}=y_k=w=0$ and 
\beqa
z_k({\textstyle \frac14}b_{2,1}
+{\textstyle \frac12}b_{4,k+1}z_k
+{\textstyle \frac14}b_{6,2k+1}z_k^2)=0.
\label{threeconifolds}
\eeqa
They are shown in figure \ref{fig:splitandover-splitI2k+1} 
as $v_{q_k}$ (when $z_k=0$), $v_{p_k}$ and $v_{r_k}$ 
(when $z_k$ is one of the roots of 
${\textstyle \frac14}b_{2,1}
+{\textstyle \frac12}b_{4,k+1}z_k
+{\textstyle \frac14}b_{6,2k+1}z_k^2=0$). 
In the split case, the two points where $z_k$ is a non-zero 
root of the latter equation are not conifold singularities
since
the second term in (\ref{I2k+1splitconifoldeq}) is 
$O(w^0)$ near these points, whereas
in the non-split case, the second term 
in (\ref{I2k+1oversplitconifoldeq}) is $O(w^2)$ there.

We can see that, unlike the (ordinary) split $I_{2k+1}$ case, 
the equation (\ref{I2k+1oversplitconifoldeq}) is a function of $w^2$,
so we can do the same unfolding $w^2\rightarrow w^2-\epsilon^2$ 
as we did in the $I_{2k}$ models. Again, on one hand, 
this replacement amounts to 
deforming all the conifold singularities occurring at $w=0$,
and on the other hand, one of the square factors of 
$b_{2,0}$ becomes generic, which turns the over-split $I_{2k+1}$ model
into a non-split $I_{2k+1}$ model.

\subsection{The split/non-split transitions and conifold transitions 
in the $I_3$ models}
Finally, to make the discussion complete, let us briefly describe the 
split/non-split transitions in the $I_{2k+1}$ models for $k=1$, i.e. the $I_3$ 
model. This lowest $k$ case is rather special and exhibits 
slightly different intersection patterns of the exceptional curves.

\begin{figure}[ht]
   \centering
   \includegraphics[width=180pt]{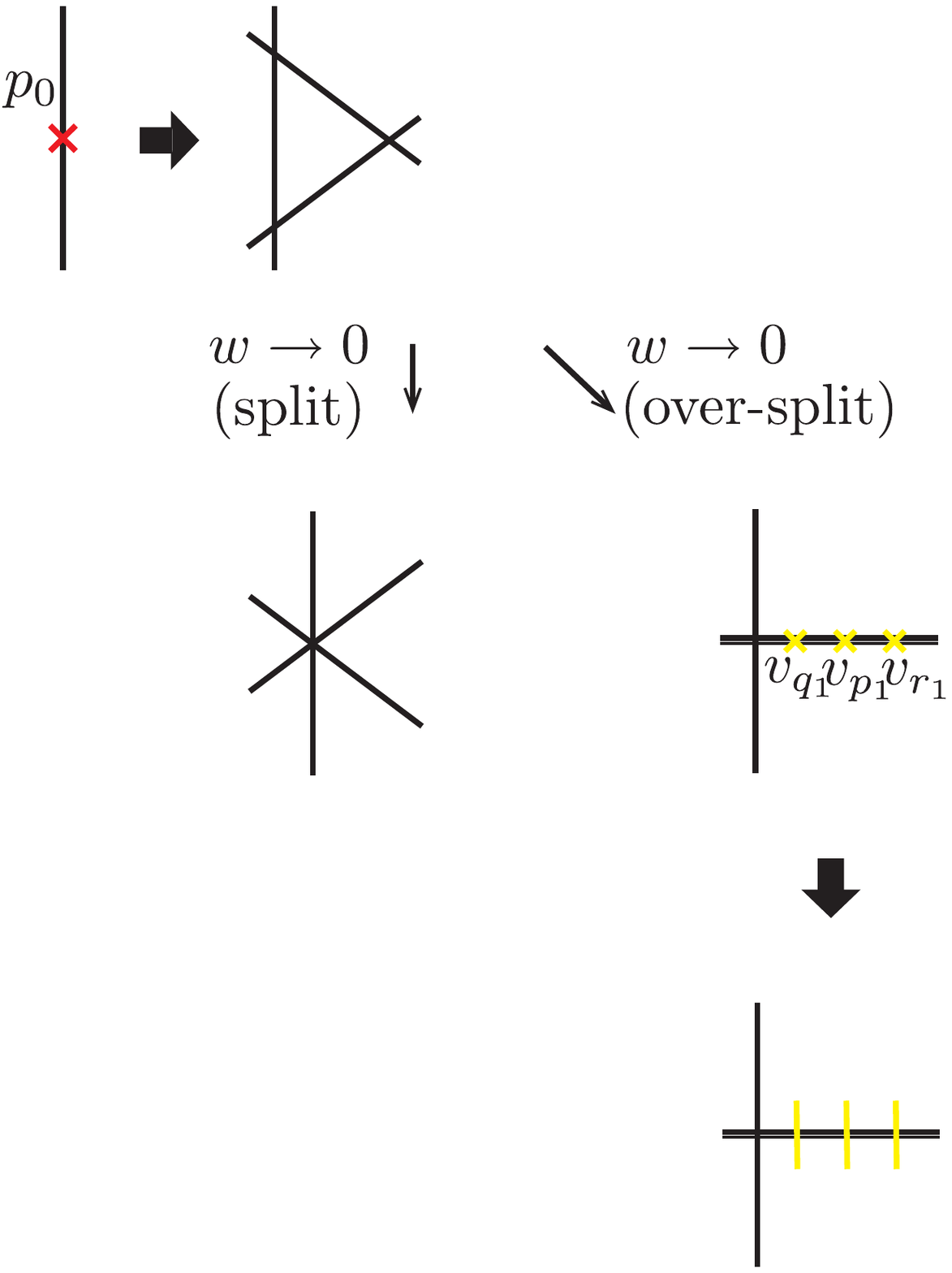} 
  \caption{Singularities and exceptional curves 
  in a split and an over-split $I_{3}$ model. }
   \label{fig:I3os}
\end{figure}

We have shown in figure \ref{fig:I3os} the 
singularities and exceptional curves 
  in a split and an over-split $I_{3}$ model near a double root of $c_{1,0}^2=0$. 
In an ordinary split $I_3$ model, no conifold singularity appears 
once the ``codimension-one'' singularity is blown up, even 
when 
$c_{1,0}\equiv w$ 
is taken to zero, where the fiber type changes from $I_3$ to $IV$. No 
matter hypermultiplet arises at the zero loci of $c_{1,0}$. In 
the over-split $I_3$ model, where we take
\beqa
b_{2,0}&=&c_{1,0}^2\n
b_{4,1}&=&c_{1,0}^2 c_{2,1},\n
b_{6,2}&=&c_{1,0}^2 c_{2,1}^2,
\label{I3oversplitb2b4b6}
\eeqa
three conifold singularities appear at each zero locus of $c_{1,0}$,
whose small resolutions yield exceptional curves of the $I_0^*$ type, 
and the singularity is enhanced from $A_2$ to $D_4$. 

Although the way the conifold singularities appear is slightly different 
from the cases for $k\geq 2$, 
the over-split $I_3$ model is also turned into the non-split $I_3$ model
by the replacement $w^2\rightarrow w^2-\epsilon^2$, which is 
a deformation of a conifold singularity. 
\section{Split/non-split transitions as conifold transitions (III): $IV$}
Let us next consider the $IV$ model. 
The $IV$ model is defined in the Deligne form (\ref{Deligne}) 
for $b_2$, $b_4$, $b_6$ with vanishing orders $1$, $2$, $2$, 
respectively.
The sections $f$, $g$ characterizing the Weierstrass equation 
read
\beqa
f&=&-\frac1{48}
(b_{2,1}^2-24 b_{4,2}) z^2
+\cdots,\n
g&=&\frac{1}{4}b_{6,2}z^2
+\cdots,\n
\label{fandgsplitIV}
\eeqa
and the discriminant is
\beqa
\Delta&=&\frac{27}{16} b_{6,2}^2 z^4+\cdots,
\label{DeltasplitIV}
\eeqa
so ord$(f,g,\Delta)=(2,2,4)$ and the generic 
fiber type at $z=0$ is $IV$. 
At the zero loci of $b_{6,2}$, 
they are enhanced to $(2,3,6)$,  showing that the Kodaira 
fiber type there is $I_0^*$. 
If the section $b_{6,2}$ can 
 be written in the form of a square $c_{3,1}^2$
for some $c_{3,1}$, the model is said a split $IV$ model, 
while if $b_{6,2}$ cannot be written that way,  it is said 
a non-split $IV$ model \cite{BIKMSV}.

In this case, the only  ``codimension-one'' singularity 
at a generic point on $z=0$ is 
$p_0:(x,y,z,w)=(0,0,0,w)$, which can be resolved 
by just a one-time blow-up. 
The resulting exceptional curves split into two, 
which intersect the original fiber at a single point;
they come on top of each other at $b_{6,2}=0$.

\begin{figure}[ht]
   \centering
   \includegraphics[width=165pt]{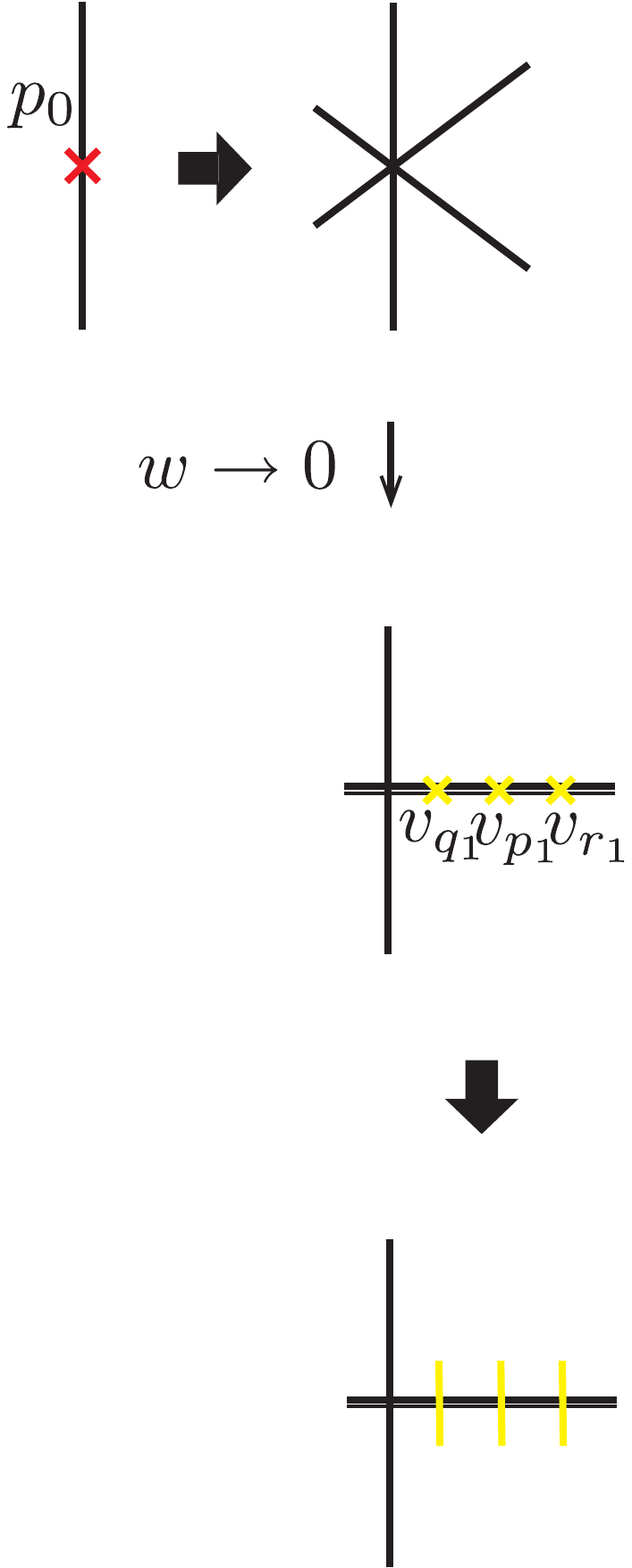} 
  \caption{Singularities and exceptional curves 
  in a split $IV$ model. }
   \label{fig:IV}
\end{figure}

In the split case, they are all double roots, 
and three new conifold singularities appear 
on the overlapping exceptional lines.
To see this, consider the 
equation blown up once $\Phi_z(x_1, y_1, z,w)=0$ with
\beqa
\Phi_z(x_1, y_1, z ,w)
&=&
-y_1^2+x_1^3 z \n
&&
+{\textstyle \frac14}(b_{2,1}z+\cdots)x_1^2
+{\textstyle \frac12}(b_{4,2}z+\cdots)x_1
+{\textstyle \frac14}(w^2+b_{6,3}z+\cdots)\n
&\stackrel{z\rightarrow 0}{\rightarrow}& 
 -y_1^2
+{\textstyle \frac14} w^2,
\label{PhizIV}
\eeqa
in ${\bf 1}_z$, where we have set $b_{6,2}=w^2$ 
to focus on a particular double root of $b_{6,2}=0$.
(\ref{PhizIV}) indeed shows that the generic exceptional curve 
splits into two lines, and they coincide with each other at $w=0$.
Conifold singularities can be seen by rewriting (\ref{PhizIV}) as
\beqa
\Phi_z(x_1, y_1, z ,w)
&=&
-y_1^2+{\textstyle \frac14} w^2+z(x_1^3  
+{\textstyle \frac14} b_{2,1} x_1^2
+{\textstyle \frac12} b_{4,2} x_1
+{\textstyle \frac14} b_{6,3} + O(z)).
\label{PhizIVconifoldeq}
\eeqa
For generic $b_{2,1}$, $b_{4,2}$, $b_{6,3}$, 
the cubic equation of $x_1$ has three distinct roots, 
giving rise to three conifold singularities at $y_1=w=z=0$.
Again, the replacement $w^2\rightarrow w^2-\epsilon^2$ 
amounts to the transition from the split to non-split $IV$ model,
at the same time it unfolds the conifold singularity to 
yield a local deformed conifold. 
Singularities and exceptional curves in the split $IV$
 model near $w=0$ are depicted in figure \ref{fig:IV}. 

\section{Split/non-split transitions as conifold transitions (IV): $IV^*$  }
In the $IV^*$ model, the vanishing orders of 
$b_2, b_4, b_6$ are $2, 3, 4$, respectively. 
$f$ and $g$ (\ref{fandg}) are
\beqa
f&=&\frac1{2}b_{4,3}z^3+\cdots,\n
g&=&\frac14 b_{6,4}z^4+\cdots.
\eeqa
The discriminant is 
\beqa
\Delta&=&\frac{27}{16}b_{6,4}^2 z^8 +\cdots.
\eeqa
These imply that the fiber type is $IV^*$ at a generic point of $z=0$.
The split $IV^*$ model has $b_{6,4}$ in the form of a square $c_{3,2}^2$ 
for some $c_{3,2}$. The non-split $IV^*$ model has generic $b_{6,4}$
\cite{BIKMSV}. In both the split and non-split models, 
the vanishing orders of $(f, g, \Delta)$ at the zero locus of $b_{6,4}$ 
changes from $(3,4,8)$ to $(3,5,9)$, 
implying that the apparent fiber type there is $III^*$, that is,
the zero locus of $b_{6,4}$ is an $E_7$ point.
  
\begin{figure}[ht]
   \centering
   \includegraphics[width=280pt]{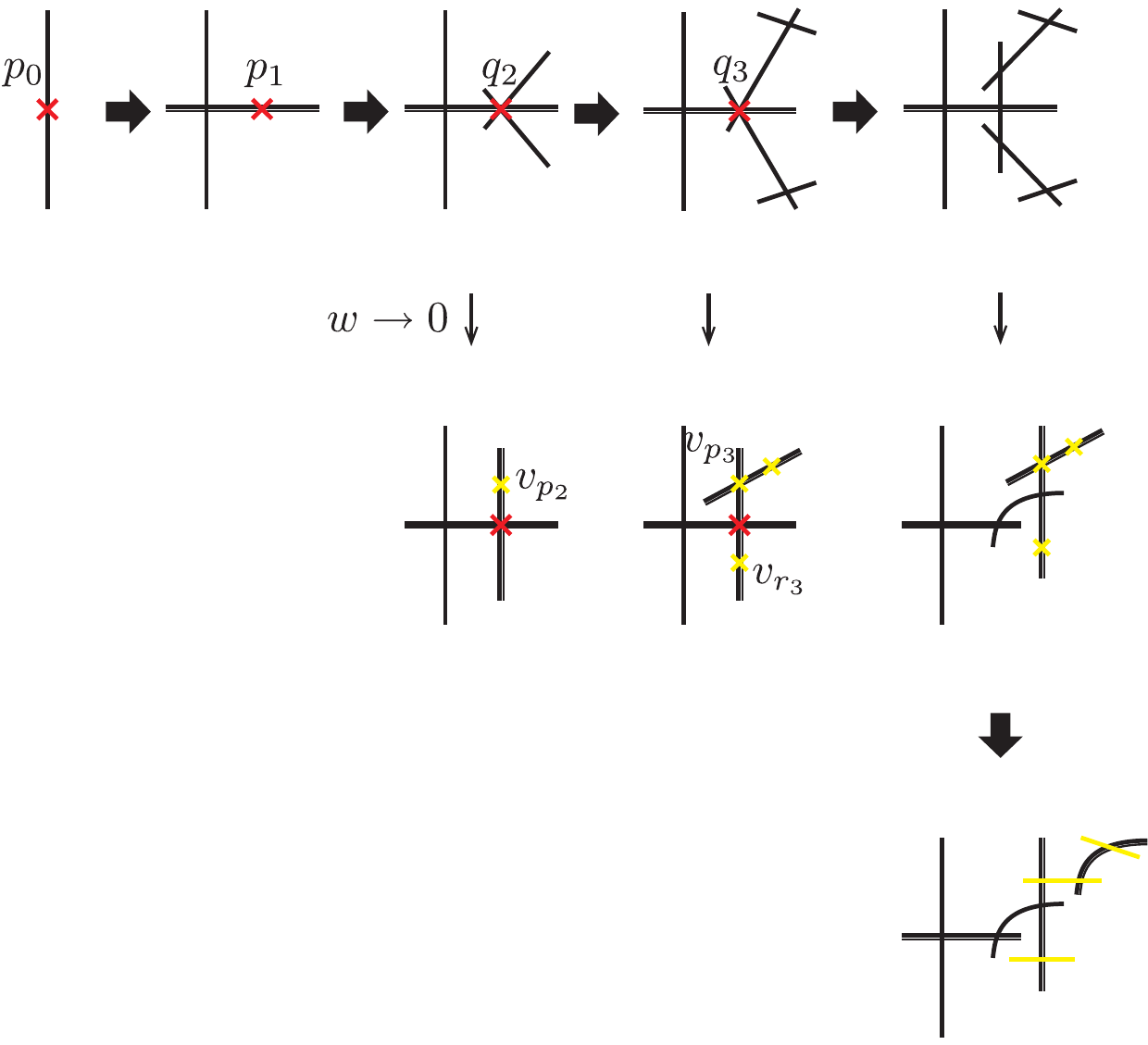} 
  \caption{Singularities and exceptional curves 
  in the split  $IV^*$ model 
  near a double root of $c_{3,2}^2=0$. }
   \label{fig:IV*}
\end{figure}

We have illustrated in figure \ref{fig:IV*} 
how the singularities appear and exceptional curves 
intersect 
in the split $IV^*$ model near $w=0$, which 
is one of the double roots of $c_{3,2}^2=0$.
At the stage where the three ``codimension-one'' singularities 
are blown up, there remain three conifold singularities 
at each double root of  $b_{6,4}=c_{3,2}^2=0$.
We will show that, if all these conifold singularities are 
resolved by small resolutions, we obtain a smooth, fully 
resolved split $IV^*$ model, while if all the conifold singularities 
are simultaneously deformed, we are led to a smooth 
non-split $IV^*$ model.

We start with a split $IV^*$ model.
The defining equation is\footnote{Although we are interested in
the local structure of the singularity, the $IV^*$ models are 
well-defined as a ${\rm dP}_9$ fibration to consider the heterotic dual,
so we have kept in (\ref{IV*Phi}) 
only terms with coefficients $b_{k,j}$ up to $j\leq k$. 
In any case, it doesn't really matter whether we do so or not.
} 
\beqa
\Phi(x,y,z,w)\equiv&-y^2+x^3 &+{\textstyle \frac14} b_{2,2}z^2 x^2 \n
&&
+ {\textstyle \frac12}  (b_{4,3}z^3+b_{4,4}z^4)x 
\n&&
+{\textstyle \frac14} (c_{3,2}^2z^{4}+b_{6,5}z^{5}+\cdots)=0.
\label{IV*Phi}
\eeqa
The first ``codimension-one'' singularity 
(next to the original singularity $p_0$) can be found 
on $\Phi_z(x_1, y_1, z,w)=0$ 
defined in (\ref{In1stblowup}) with $\Phi(x,y,z,w)$ given by 
(\ref{IV*Phi}). This is
\beqa
p_1:&&(x_1,y_1,z,w)=(0,0,0,0)~~~({\mbox{\boldmath $1$}}_{z}).
\eeqa
Blowing up $\Phi_z(x_1, y_1, z,w)=0$ at $p_1$, we have
\beqa
\Phi_{zx}(x_1, y_2, z_2,w)=&-y_2^2+x_1^2 z_2 &
+{\textstyle \frac14} b_{2,2} x_1^2 z_2^2 \n
&&
+ {\textstyle \frac12}  (b_{4,3} x_1 z_2^2+b_{4,4}x_1^2 z_2^3)
\n&&
+{\textstyle \frac14} (c_{3,2}^2 z_2^{2}+b_{6,5} x_1 z_2^{3}+\cdots)=0,
\label{IV*Phizx}
\eeqa
where  $\Phi_{zx}(x_1, y_2, z_2,w)$
is defined similarly to (\ref{In2ndblowup}).
In the $x_1\rightarrow 0$ limit, this equation reduces to
$y_2^2=0$, which is a double line. It has a ``codimension-one'' singularity
\beqa
q_2:&&(x_1,y_2,z_2,w)=(0,0,0,w)~~~({\mbox{\boldmath $2$}}_{zx})
\eeqa
as well as a conifold singularity
\beqa
v_{p_2}:&&(x_1,y_2,z_2,w)=(0,0,-\frac{2 b_{4,3}}{b_{6,5}},0)~~~({\mbox{\boldmath $2$}}_{zx}).
\eeqa
The latter can be seen by writing (\ref{IV*Phizx}) as
\beqa
-y_2^2+{\textstyle \frac14} w^2 z_2^{2} 
+  x_1 ({\textstyle \frac12} b_{4,3} z_2^2 + {\textstyle \frac14}b_{6,5}  z_2^{3}+O(x_1))=0, 
\label{IV*Phizxconifoldeq}
\eeqa
where we again set $c_{3,2}^2=w^2$ to focus on a particular double root 
of  $b_{6,4}=c_{3,2}^2=0$.

Blowing up $\Phi_{zx}(x_1, y_2, z_2,w)=0$ at $q_2$, 
we have 
\beqa
\Phi_{zxx}(x_1, y_3, z_3,w)=&-y_3^2+x_1 z_3 &
+{\textstyle \frac14} b_{2,2} x_1^2 z_3^2 \n
&&
+ {\textstyle \frac12}  (b_{4,3} x_1 z_3^2+b_{4,4}x_1^3 z_3^3)
\n&&
+{\textstyle \frac14} (c_{3,2}^2 z_3^{2}+b_{6,5} x_1^2 z_3^{3}+\cdots)=0 
\label{IV*Phizxx}
\eeqa
in the patch ${\mbox{\boldmath $3$}}_{zxx}$,
where we have defined
\beqa
\Phi_{zxx}(x_1, y_3, z_3,w)&\equiv&
x_1^{-2}
\Phi_{zx}(x_1, x_1 y_3, x_1 z_3,w).
\eeqa
(\ref{IV*Phizxx}) still has a ``codimension-one'' singularity
\beqa
q_3:&&(x_1,y_3,z_3,w)=(0,0,0,w)~~~({\mbox{\boldmath $3$}}_{zxx}).
\eeqa
(\ref{IV*Phizxx}) has also a conifold equation, but in fact, there arise 
two conifold singularities after blowing up at $q_2$ as we displayed 
in figure \ref{fig:IV*}, and it is only the one of two that can be seen in 
the patch ${\mbox{\boldmath $3$}}_{zxx}$. 

To see both conifold singularities we consider 
\beqa
\Phi_{zxz}(x_3, y_3, z_2,w)=&-y_3^2+x_3^2 z_2 &
+{\textstyle \frac14} b_{2,2} x_3^2 z_2^2 \n
&&
+ {\textstyle \frac12}  (b_{4,3} x_3 z_2+b_{4,4}x_3^2 z_2^3)
\n&&
+{\textstyle \frac14} (c_{3,2}^2 +b_{6,5} x_3 z_2^{2}+\cdots)=0
\label{IV*Phizxz}
\eeqa
in the patch ${\mbox{\boldmath $3$}}_{zxz}$,
where 
\beqa
\Phi_{zxz}(x_3, y_3, z_2,w)&\equiv&
z_2^{-2}
\Phi_{zx}(x_3 z_2, y_3 z_2, z_2,w).
\eeqa
(\ref{IV*Phizxz}) can also be transformed into the form of a conifold equation 
\beqa
-y_3^2+{\textstyle \frac14} w^2 +z_2\left(x_3^2+ {\textstyle \frac12} b_{4,3}x_3+O(z_2)
\right)
=0,
\label{IV*Phizxzconifoldeq}
\eeqa
which indicates the existence of two conifold singularities
\beqa
v_{p_3}:&&(x_3,y_3,z_2,w)
=(0,0,0,0),\n
v_{r_3}:&&(x_3,y_3,z_2,w)
=(-{\textstyle \frac12} b_{4,3},0,0,0)~~~({\mbox{\boldmath $3$}}_{zxz}).
\eeqa

By looking at the form of the conifold equations 
 (\ref{IV*Phizxconifoldeq})
and
 (\ref{IV*Phizxzconifoldeq}) and following the discussion we have 
 presented in the previous sections, it is now clear that 
 the transition from the split $IV^*$ model to the 
 non-split $IV^*$ model is the conifold transition from 
 the resolved side to the deformed side. 
 Note that this is the only example in which the transition 
 occurs at an $E_7$ point; as we saw in the previous sections, 
 as well as we will see in the next section, the transition always 
 occurs at a $D_{2k}$ point in all the other examples.

\section{The $I_{n}^*$ models}
Finally, we will deal with the $I_{n}^*$ cases. The situation is quite different 
when $n$ is even and when $n$ is odd. We will consider the odd case first.

\subsection{The $I_{2k-3}^*$ models}
The  $I_{2k-3}^*$ models $(k\geq 2)$ have a $D_{2k+1}$ singularity.
In the split $I_{2k-3}^*$ models $(k\geq 2)$, conifold singularities 
appear as in the previous examples, and the deformation 
at the $D_{2k+2}$ points turns a split $I_{2k-3}^*$ model into 
a non-split one and can be regarded as a deformation of the 
conifold singularities.

\begin{figure}[ht]
   \centering
   \includegraphics[width=435pt]{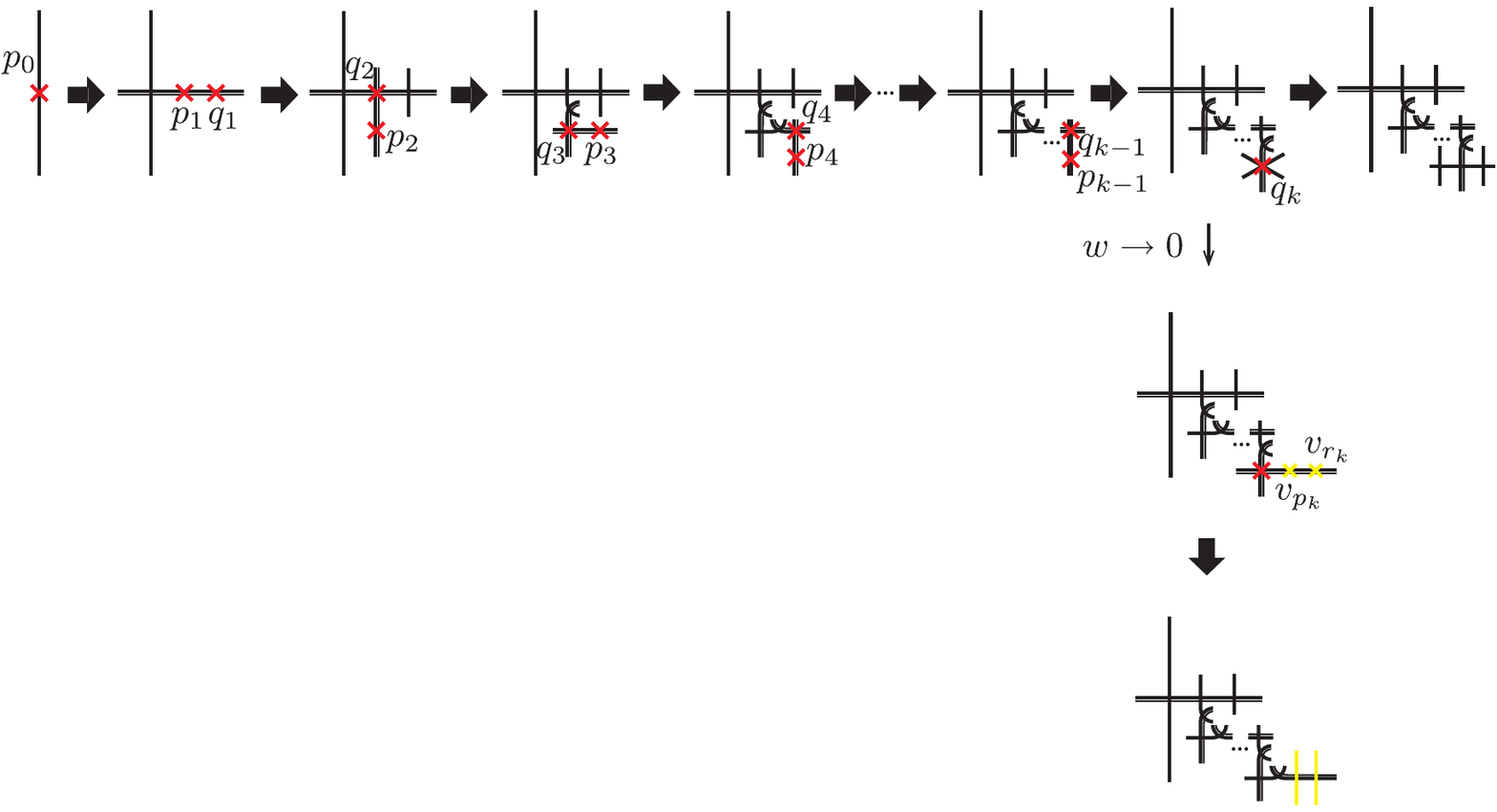} 
  \caption{Singularities and exceptional curves 
  in the split  $I_{2k-3}^*$ model. }
   \label{fig:I2k-3star}
\end{figure}

The model is defined by (\ref{Deligne}) with vanishing orders 
$\mbox{ord}(b_2,b_4,b_6)=(1,k+1,2k)$ ($k\geq 2$). 
Whether the model is split or non-split depends on whether or not 
the section $b_{6,2k}$ takes the form of a square $c_{3,k}^2$ for some
$c_{3,k}$ \cite{BIKMSV}. In the split case, the Lie algebra of the 
unbroken gauge symmetry is $D_{2k+1}=SO(4k+2)$. Whether split or non-split,
the zero loci of $b_{6,2k}$ are $D_{2k+2}=SO(4k+4)$ points. 
Besides them, $E_6$ and $E_8$ points may occur for $k=2$ and $3$, 
but they are not important here.

As we have shown in figure \ref{fig:I2k-3star}, one of the differences in the 
split $I_n^*$ model is that the conifold singularities appear only at the final 
step of blowing up. 
We can see the conifold singularities in the equation 
$\Phi_{\underbrace{\mbox{\scriptsize $z\cdots z$}}_k}(x_k,y_k,z,w)=0$,
where, setting $c_{3,k}^2\equiv w^2$,
\beqa
\Phi_{\underbrace{\mbox{\scriptsize $z\cdots z$}}_k}(x_k,y_k,z,w)=
&-y_k^2+x_k^3 z^k &
+{\textstyle \frac14} (b_{2,1} z+\cdots)x_k^2 \n
&&
+ {\textstyle \frac12}  (b_{4,k+1} z+\cdots)x_k
\n&&
+{\textstyle \frac14} (w^2 +b_{6,2k+1} z+\cdots)
\n
=&-y_k^2+{\textstyle \frac14} w^2&+
z\left(
{\textstyle \frac14} b_{2,1}x_k^2+ {\textstyle \frac12}  b_{4,k+1} x_k
+{\textstyle \frac14} b_{6,2k+1}
+O(z)
\right). 
\label{I2k-3*Phizk}
\eeqa
The discriminant of the quadratic equation ${\textstyle \frac14} b_{2,1}x_k^2+ {\textstyle \frac12}  b_{4,k+1} x_k
+{\textstyle \frac14} b_{6,2k+1}=0$ is proportional to $b_{8,2k+2}$, 
which does not vanish generically. Therefore it has two distinct roots, 
yielding the two conifold singularities. The equation (\ref{I2k-3*Phizk}) 
again depends on $w$ through $w^2$ near the singularities,
and unfolding the conifold singularity is exactly what 
turns a split model into a non-split one.

\subsection{The $I_{2k-2}^*$ models}
So far we have seen various examples in which the split/non-split transition 
is precisely the conifold transition associated with the conifold singularities 
occurring at the $D_{2k}$ points, or the $E_7$ points in the $IV^*$ case.
In fact, in the $I_{2k-2}^*$ model, the situation is quite different.
The crucial difference is that, in that case, no conifold singularity 
arises at the zero locus of the section relevant to the split/non-split 
transition.

In this class of models, the orders of $b_2$, $b_4$, $b_6$ are 
$1$, $k+1$, $2k+1$, instead of $1$, $k+1$, $2k$ in the previous $I_{2k-3}^*$ 
models. $k=1$ is a special case and has already been discussed 
in detail in \cite{EsoleetalSO(8)}\footnote{
For the $I_0^*$ models, 
 we have, again, presented in table \ref{split/nonsplit} 
 the {\em generic} orders of $b_2$, $b_4$, $b_6$ ($=$ $1$, $2$, $3$) 
 that can achieve these fiber types with the additional constraints 
 shown there. For the split and semi-split $I_0^*$ models, 
 $p_{2,1}$ can be eliminated by a redefinition of $x$, 
 so that the orders of $b_2$, $b_4$, $b_6$ become $1$, $2$, $4$, 
which are the values derived from the standard Tate's orders for 
  the split and semi-split $I_0^*$ models.
 }, so we will consider $k\geq 2$.
$f$ and $g$ (\ref{fandg}) read 
\beqa
f&=&-\frac1{48} b_{2,1}^2 z^2 +\cdots,\n
g&=&+\frac1{864} b_{2,1}^3 z^3 +\cdots,
\eeqa
which are the same as those in the $I_{2k-3}^*$ models.
The discriminant is 
\beqa
\Delta&=&\frac1{16}b_{2,1}^2 b_{8,2k+2} z^{2k+4} +\cdots,
\eeqa
so, for a generic $b_{2,1}$, the singularity is enhanced from 
$D_{2k+2}$ to $D_{2k+3}$ at the zero locus of $b_{8,2k+2}$,
where 
\beqa
b_{8,2k+2}&=&\frac14(b_{2,1}b_{6,2k+1}-b_{4,k+1}^2).
\eeqa
If this $b_{8,2k+2}$ is written as $c_{4,k+1}^2$ for some $c_{4,k+1}$,
this $I_{2k-2}^*$ model is called split, otherwise non-split \cite{BIKMSV}. 

\begin{figure}[ht]
   \centering
   \includegraphics[width=310pt]{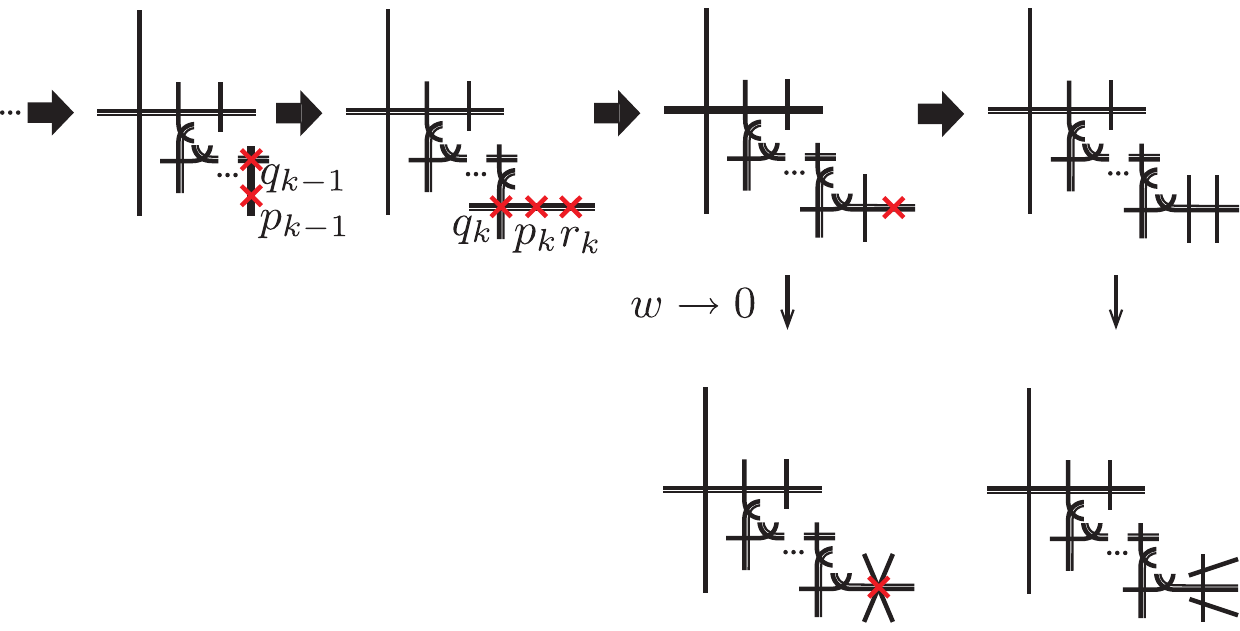} 
  \caption{Singularities and exceptional curves 
  in the split  $I_{2k-2}^*$ model. }
   \label{fig:I2k-2star}
\end{figure}

The blowing-up procedure proceeds similarly to 
the $I_{2k-3}^*$ models. In the split case, 
a difference arises when $p_{k-1}$ is blown up, 
where the exceptional curves overlap to one line 
instead of splitting into two lines, and 
three ``codimension-one'' singularities arise on the line.
This is precisely what was seen in the $w\rightarrow 0$ limit 
after $p_{k-1}$ was blown up in the $I_{2k-3}^*$ models, 
where the two conifold singularities found there are now 
replaced by two ``codimension-one'' singularities (figure \ref{fig:I2k-2star}).
Concretely, 
\beqa
\Phi_{\underbrace{\mbox{\scriptsize $z\cdots z$}}_k}(x_k,y_k,z,w)=
&-y_k^2+x_k^3 z^k &
+{\textstyle \frac14} (b_{2,1} z+\cdots)x_k^2 \n
&&
+ {\textstyle \frac12}  (b_{4,k+1} z+\cdots)x_k
\n&&
+{\textstyle \frac14} (b_{6,2k+1} z+\cdots).
\label{I2k-2*Phizk}
\eeqa
Since $b_{8,2k+2}$ is proportional to the 
discriminant of the quadratic equation of 
${\textstyle \frac14} b_{2,1}x_k^2+ {\textstyle \frac12}  b_{4,k+1} x_k
+{\textstyle \frac14} b_{6,2k+1}=0$, we can further 
write, by assuming $b_{8,2k+2}=c_{4,k+1}^2$, as
\beqa
\Phi_{\underbrace{\mbox{\scriptsize $z\cdots z$}}_k}(x_k,y_k,z,w)&=&
-y_k^2+
z\left(
{\textstyle \frac14} b_{2,1}x_k^2+ {\textstyle \frac12}  b_{4,k+1} x_k
+{\textstyle \frac14} b_{6,2k+1}
+O(z)
\right)\n
&=&
-y_k^2+
\frac{z}{b_{2,1}}\left(
\left(
\frac{b_{2,1}}2 x_k + b_{4,k+1}
\right)^2 +c_{4,k+1}^2
+O(z)
\right).
\label{I2k-2*Phizksplit}
\eeqa
Thus, the ``codimension-one'' singular loci of 
$\Phi_{\underbrace{\mbox{\scriptsize $z\cdots z$}}_k}(x_k,y_k,z,w)=0$
split into two irreducible components 
\beqa
y_k=0, ~z=0,~\frac{b_{2,1}}2 x_k + b_{4,k+1} \pm i c_{4,k+1}=0.
\label{twocomponents}
\eeqa
Their intersection is where $c_{4,k+1}$ vanishes, or 
equivalently, $b_{8,2k+2}=0$ vanishes, so it is a $D_{2k+3}$ point.
The ``codimension-one'' singularities can be blown up along 
either of the two irreducible components (\ref{twocomponents}) first.
One can verify that the exceptional curve obtained in such a way 
splits into two lines precisely at the intersection $D_{2k+3}$ point.
Blowing up along the remaining irreducible component thus yields 
the $D_{2k+3}$ intersection diagram only there. 
This is how the higher-rank intersection diagram emerges 
without conifold singularities in the $I_{2k-2}^*$ models.

On the other hand, the equation of the non-split $I_{2k-2}^*$ model 
can be obtained by replacing $c_{4,k+1}^2$ with a generic
$b_{8,2k+2}$ in (\ref{I2k-2*Phizksplit}). In this case, 
the ``codimension-one'' singular loci consist of 
only one irreducible component, along which we can blow up 
the singularities only once. No conifold singularity is found. 
Therefore, only the $I_{2k-2}^*$ models 
(including the $I_{0}^*$ model \cite{EsoleetalSO(8)}) 
cannot interpret the split/non-split transition there as a conifold transition. 

\section{Conclusions}
In this paper, we have shown that in six-dimensional 
 F-theory on elliptic Calabi-Yau threefolds on the Hirzebruch 
 surface $\FF_n$, 
 all the non-split models listed in \cite{BIKMSV},  except a certain class of fiber types, can be realized by a conifold transition  from the corresponding split models.
We examined this fact separately for all cases 
of $I_n$ ($n\geq 3$), $I_n^*$ ($n\geq 0$), $IV$ and $IV^*$, 
in which there is a distinction between the 
split and non-split types.

In the split models of the fiber types $I_{2k}$ ($k\geq 2$), $IV$ and 
$I_{2k-3}^*$ ($k\geq 2$), there generically exist points where 
the singularities $SU(2k)$, $SU(3)$ and $SO(4k+2)$ are enhanced 
to $SO(4k)$, $SO(8)$ and $SO(4k+4)$ in the sense of Kodaira. 
When all the ``codimension-one'' singularities are blown up, 
there remain some conifold singularities there. If these conifold 
singularities are resolved by small resolutions, one obtains a smooth 
split model for each case. This is the resolved side of the conifold 
transition. On the other hand, at the stage where all the 
``codimension-one'' singularities are blown up, 
one can also deform the relevant section so that the 
split model transforms into the non-split model, thereby 
all the conifold singularities are simultaneously unfolded. 
This is the deformed side of the conifold transition.

The $IV^*$ model is similar to these models, but 
only in this case, the singularity of the enhanced point
where the conifold transition occurs is $E_7$ instead of $SO(4k)$.

The split $I_{2k-1}$ model has generically an $SU(2k-1)$ singularity and 
has no $SO(4k)$ point in general. However, by adjusting the complex 
structure, one can make the $SO(4k-2)$ point and the $SU(2k)$ point 
come to the same point to achieve an $SO(4k)$ point. 
We called such a split $I_{2k-1}$ model  with a special complex 
structure an ``over-split'' model. We found that in this case, 
there arose conifold singularities at the $SO(4k)=D_{2k}$ point 
after blowing up all the ``codimension-one'' singularities.
Then the non-split $I_{2k-1}$ model is obtained by the 
deformation similarly.

Finally, in the case of the $I_{2k-2}^*$ models, 
the conifold singularity does not appear after the 
blow-up of the ``codimension-one'' singularities.
Therefore, this is a special case in which 
the split/non-split transition cannot be regarded 
as a conifold transition.

We have also examined how the proposal of \cite{AKM} for
the origin of non-local matter can be actually 
implemented in our blow-up analysis. 
We have shown that the genus-$g$ Riemann surface that 
plays the essential role in the proposal can be obtained as 
an intersection of the blown-up three-fold and an appropriate 
divisor, and by forgetting the $\PP^1$ fiber direction.  
It has also been found that even when there are multiple pairs 
of exceptional curves, the genus-$g$ Riemann surface obtained 
from them are all identical and thus well-defined.
%

Conifold singularities are ubiquitous, associated with 
matter generations in F-theory. 
As we stressed, 
these are not the ones created by some fine tuning 
of moduli, but always occur where matter is generated 
in the very general setting in F-theory.
The conifold transition has been an important 
key concept in discussions in AdS/CFT \cite{KW,KS}, 
topological string theory \cite{GV,VafalargeN}, 
and string cosmology (e.g. \cite{GKP,KKLT}).
In view of these facts, it would be very interesting to consider 
new applications of the facts revealed here to the theory of 
superstring phenomenology and cosmology.

%
%

\acknowledgments

We thank H.~Itoyama, Y.~Kimura and H.~Otsuka 
for valuable discussions.



\begin{thebibliography}{99}

%
%



\bibitem{Vafa}
C.~Vafa,
  \emph{Evidence for F theory},
  Nucl.\ Phys.\ B {\bf 469}, 403 (1996), arXiv: hep-th/9602022.


\bibitem{MV1}
D.~R.~Morrison and C.~Vafa,
\emph{Compactifications of F theory on Calabi-Yau threefolds. 1},
  Nucl.\ Phys.\ B {\bf 473}, 74 (1996), arXiv: hep-th/9602114.

\bibitem{MV2}
D.~R.~Morrison and C.~Vafa,
  \emph{Compactifications of F theory on Calabi-Yau threefolds. 2.},
  Nucl.\ Phys.\ B {\bf 476}, 437 (1996), arXiv: hep-th/9603161.

\bibitem{BIKMSV} M.~Bershadsky, K.~Intriligator, S.~Kachru, D.R.~Morrison, V.~Sadov and C.~Vafa,
\emph{Geometric singularities and enhanced gauge symmetries},
Nucl.Phys. B481 (1996) 215-252, arXiv: hep-th/9605200.

\bibitem{KatzVafa}
S.~H.~Katz and C.~Vafa,
  \emph{Matter from geometry},
  Nucl.\ Phys.\ B {\bf 497}, 146 (1997), arXiv: hep-th/9606086.
  
\bibitem{Kodaira}
K.~Kodaira, Ann. of  Math. {\bf 77}, 563 (1963).

\bibitem{DonagiWijnholt}
R.~Donagi and M.~Wijnholt,
\emph{Model Building with F-Theory},
  Adv.\ Theor.\ Math.\ Phys.\  {\bf 15}, 1237 (2011), arXiv:0802.2969 [hep-th].



  \bibitem{BHV}
  C.~Beasley, J.~J.~Heckman and C.~Vafa,
\emph{GUTs and Exceptionl Branes in F-theory - I},
  JHEP {\bf 0901}, 058 (2009), arXiv:0802.3391 [hep-th].


    \bibitem{BHV2}
  C.~Beasley, J.~J.~Heckman and C.~Vafa,
\emph{GUTs and Exceptional Branes in F-theory - II: Experimental Predictions},
  JHEP {\bf 0901}, 059 (2009), arXiv:0806.0102 [hep-th].

 
  \bibitem{DonagiWijnholt2}
  R.~Donagi and M.~Wijnholt,
 \emph{Breaking GUT Groups in F-Theory},
  Adv.\ Theor.\ Math.\ Phys.\  {\bf 15}, 1523 (2011), arXiv:0808.2223 [hep-th].


\bibitem{HKTW}
  H.~Hayashi, T.~Kawano, R.~Tatar and T.~Watari,
\emph{Codimension-3 Singularities and Yukawa Couplings in F-theory},
  Nucl.\ Phys.\ B {\bf 823} (2009) 47
,arXiv:0901.4941 [hep-th].

\bibitem{DWHiggsBundles}
  R.~Donagi and M.~Wijnholt,
\emph{Higgs Bundles and UV Completion in F-Theory},
  Commun.\ Math.\ Phys.\  {\bf 326} (2014) 287
,arXiv:0904.1218 [hep-th].



\bibitem{localmodel1}
J.~J.~Heckman, J.~Marsano, N.~Saulina, S.~Schafer-Nameki and C.~Vafa,
\emph{Instantons and SUSY breaking in F-theory},
arXiv:0808.1286 [hep-th].
 %
\bibitem{localmodel2}
J.~Marsano, N.~Saulina and S.~Schafer-Nameki,
\emph{Gauge Mediation in F-Theory GUT Models},
  Phys.\ Rev.\ D {\bf 80} (2009) 046006
, arXiv:0808.1571 [hep-th].
  %
  \bibitem{localmodel3}
 J.~J.~Heckman and C.~Vafa,
\emph{F-theory, GUTs, and the Weak Scale},
  JHEP {\bf 0909} (2009) 079
, arXiv:0809.1098 [hep-th].
%
\bibitem{localmodel4}
A.~Font and L.~E.~Ibanez,
\emph{Yukawa Structure from U(1) Fluxes in F-theory Grand Unification},
  JHEP {\bf 0902} (2009) 016
, arXiv:0811.2157 [hep-th].
 %


\bibitem{MT}
  D.~R.~Morrison and W.~Taylor,
  \emph{Matter and singularities},
  JHEP {\bf 1201}, 022 (2012),
arXiv:1106.3563 [hep-th].

\bibitem{EsoleYau} 
  M.~Esole and S.~T.~Yau,
  \emph{Small resolutions of SU(5)-models in F-theory},
  Adv.\ Theor.\ Math.\ Phys.\  {\bf 17}, no. 6, 1195 (2013),
arXiv:1107.0733 [hep-th].

\bibitem{MT1201.1943}
D.~R.~Morrison and W.~Taylor,
 \emph{Classifying bases for 6D F-theory models},
Central Eur. J. Phys. \textbf{10} (2012), 1072-1088
[arXiv:1201.1943 [hep-th]].


\bibitem{GSWest} 
  M.~B.~Green, J.~H.~Schwarz and P.~C.~West,
  \emph{Anomaly Free Chiral Theories in Six-Dimensions},
  Nucl.\ Phys.\ B {\bf 254}, 327 (1985).
 

\bibitem{GHLST}
A.~Grassi, J.~Halverson, C.~Long, J.~L.~Shaneson and J.~Tian,
\emph{Non-simply-laced Symmetry Algebras in F-theory on Singular Spaces,},
JHEP \textbf{09} (2018), 129
, arXiv:1805.06949 [hep-th].

\bibitem{AGW}
P.~Arras, A.~Grassi and T.~Weigand,
\emph{Terminal Singularities, Milnor Numbers, and Matter in F-theory},
J. Geom. Phys. \textbf{123} (2018), 71-97,
arXiv:1612.05646 [hep-th].

\bibitem{EJK}
M.~Esole, P.~Jefferson and M.~J.~Kang,
\emph{The Geometry of F$_4$-Models},
arXiv:1704.08251 [hep-th].



\bibitem{EK}
M.~Esole and M.~J.~Kang,
\emph{The Geometry of the SU(2)$\times$ G$_2$-model},
JHEP \textbf{02} (2019), 091,
arXiv:1805.03214 [hep-th].

\bibitem{EJ}
M.~Esole and P.~Jefferson,
\emph{USp(4)-models},
arXiv:1910.09536 [hep-th].

\bibitem{KMT}
N.~Kan, S.~Mizoguchi and T.~Tani,
\emph{Half-hypermultiplets and incomplete/complete resolutions in F-theory},
JHEP \textbf{08}, 063 (2020)
  , arXiv:2003.05563 [hep-th].

\bibitem{KuMT}
R.~Kuramochi, S.~Mizoguchi and T.~Tani,
\emph{Magic square and half-hypermultiplets in F-theory},
arXiv:2008.09272 [hep-th].


 \bibitem{Yukawas} 
  J.~Marsano and S.~Schafer-Nameki,
  \emph{Yukawas, G-flux, and Spectral Covers from Resolved Calabi-Yau's},
  JHEP {\bf 1111}, 098 (2011)
 , arXiv:1108.1794 [hep-th].

\bibitem{1402.2653}
H.~Hayashi, C.~Lawrie, D.~R.~Morrison and S.~Schafer-Nameki,
\emph{Box Graphs and Singular Fibers},
JHEP \textbf{05}, 048 (2014)
[arXiv:1402.2653 [hep-th]].


\bibitem{AKM}
P.~S.~Aspinwall, S.~H.~Katz and D.~R.~Morrison,
\emph{Lie groups, Calabi-Yau threefolds, and F theory},
Adv. Theor. Math. Phys. \textbf{4}, 95-126 (2000),
arXiv:hep-th/0002012 [hep-th].


\bibitem{CdlOGP}
P. Candelas, X. de la Ossa. P. Green and L. Parkes,
\emph{A Pair of Calabi-Yau Manifolds as an Exactly Soluble Superconformal Field Theory},
Nucl. Phys. {\bf B359} (1991) 21. 

\bibitem{CGH}
P. Candelas, P. Green and T. Hubsch, 
\emph{Rolling Among Calabi-Yau Vacua}, 
Nucl. Phys. {\bf B330} (1990) 49.   

\bibitem{CdlO}
P. Candelas and X. de la Ossa, \emph{Comments on Conifolds},
Nucl. Phys. {\bf B342} (1990) 246.

\bibitem{CdlO2}
P. Candelas and X. de la Ossa, \emph{Moduli Space of Calabi-Yau
Manifolds}, Nucl. Phys. {\bf B355} (1991) 455.

\bibitem{AGM}
P. Aspinwall, B. Greene and D. Morrison, {\bf B416} (1994) 414.

\bibitem{Strominger}
A.~Strominger,
\emph{Massless black holes and conifolds in string theory},
Nucl. Phys. B \textbf{451}, 96-108 (1995)
  , arXiv:hep-th/9504090 [hep-th].

\bibitem{GHS1} 
  A.~Grassi, J.~Halverson and J.~L.~Shaneson,
  \emph{Matter From Geometry Without Resolution},
  JHEP {\bf 1310}, 205 (2013)
    , arXiv:1306.1832 [hep-th].

\bibitem{GHS2} 
A.~Grassi, J.~Halverson and J.~L.~Shaneson,
\emph{Non-Abelian Gauge Symmetry and the Higgs Mechanism in F-theory},
Commun. Math. Phys. \textbf{336}, no.3, 1231-1257 (2015)
  , arXiv:1402.5962 [hep-th].

\bibitem{GKP}
S.~B.~Giddings, S.~Kachru and J.~Polchinski,
\emph{Hierarchies from fluxes in string compactifications},
Phys. Rev. D \textbf{66}, 106006 (2002)
  , arXiv:hep-th/0105097 [hep-th].

\bibitem{IJMMP}
K.~Intriligator, H.~Jockers, P.~Mayr, D.~R.~Morrison and M.~R.~Plesser,
\emph{Conifold Transitions in M-theory on Calabi-Yau Fourfolds with Background Fluxes},
Adv. Theor. Math. Phys. \textbf{17}, no.3, 601-699 (2013)
  , arXiv:1203.6662 [hep-th].



\bibitem{Aspinwall}
P.~S.~Aspinwall,
\emph{M theory versus F theory pictures of the heterotic string},
Adv. Theor. Math. Phys. \textbf{1}, 127-147 (1998)
  , arXiv:hep-th/9707014 [hep-th].


\bibitem{Deligne}
P. Deligne, 
Lecture Notes in Math., Vol. 476, Springer, Berlin, 1975.


\bibitem{EsoleetalSO(8)}
M.~Esole, R.~Jagadeesan and M.~J.~Kang,
\emph{The Geometry of G$_2$, Spin(7), and Spin(8)-models},
arXiv:1709.04913 [hep-th].



\bibitem{KMSNS}
S.~Katz, D.~R.~Morrison, S.~Schafer-Nameki and J.~Sully,
\emph{Tate's algorithm and F-theory},
JHEP \textbf{08}, 094 (2011)
  , arXiv:1106.3854 [hep-th].


\bibitem{1212.2949}
C.~Lawrie and S.~Sch\"afer-Nameki,
\emph{The Tate Form on Steroids: Resolution and Higher Codimension Fibers},
JHEP \textbf{04}, 061 (2013)
[arXiv:1212.2949 [hep-th]].



\bibitem{MizoguchiTaniLooijenga}
S.~Mizoguchi and T.~Tani,
\emph{Looijenga's weighted projective space, Tate's algorithm and Mordell-Weil Lattice in F-theory and heterotic string theory},
JHEP \textbf{11}, 053 (2016)
  , arXiv:1607.07280 [hep-th].

\bibitem{AGRT}
L.~B.~Anderson, J.~Gray, N.~Raghuram and W.~Taylor,
\emph{Matter in transition},
JHEP \textbf{04}, 080 (2016)
  , arXiv:1512.05791 [hep-th].


\bibitem{Sadov}
V.~Sadov,
\emph{Generalized Green-Schwarz mechanism in F theory},
  Phys.\ Lett.\ B {\bf 388}, 45 (1996)
  [hep-th/9606008].


\bibitem{Witten}
E.~Witten,
\emph{Phase transitions in M theory and F theory},
Nucl. Phys. B \textbf{471} (1996), 195-216
[arXiv:hep-th/9603150 [hep-th]].

\bibitem{KMP}
S.~H.~Katz, D.~R.~Morrison and M.~R.~Plesser,
\emph{Enhanced gauge symmetry in type II string theory},
Nucl. Phys. B \textbf{477} (1996), 105-140
[arXiv:hep-th/9601108 [hep-th]].


\bibitem{MizoguchiTaniNonCartan}
S.~Mizoguchi and T.~Tani,
\emph{Non-Cartan Mordell-Weil lattices of rational elliptic surfaces and heterotic/F-theory compactifications},
JHEP \textbf{03}, 121 (2019)
  , arXiv:1808.08001 [hep-th].

\bibitem{Tani}
T.~Tani,
  \emph{Matter from string junction},
  Nucl.\ Phys.\ B {\bf 602}, 434 (2001).


\bibitem{KW}
I.~R.~Klebanov and E.~Witten,
\emph{Superconformal field theory on three-branes at a Calabi-Yau singularity},
Nucl. Phys. B \textbf{536}, 199-218 (1998)
  , arXiv:hep-th/9807080 [hep-th].


\bibitem{KS}
I.~R.~Klebanov and M.~J.~Strassler,
\emph{Supergravity and a confining gauge theory: Duality cascades and chi SB resolution of naked singularities},
JHEP \textbf{08}, 052 (2000)
  , arXiv:hep-th/0007191 [hep-th].


\bibitem{GV}
R.~Gopakumar and C.~Vafa,
\emph{On the gauge theory / geometry correspondence},
Adv. Theor. Math. Phys. \textbf{3}, 1415-1443 (1999)
  , arXiv:hep-th/9811131 [hep-th].


\bibitem{VafalargeN}
C.~Vafa,
\emph{Superstrings and topological strings at large N},
J. Math. Phys. \textbf{42}, 2798-2817 (2001),
arXiv:hep-th/0008142 [hep-th].


\bibitem{KKLT}
S.~Kachru, R.~Kallosh, A.~D.~Linde and S.~P.~Trivedi,
\emph{De Sitter vacua in string theory},
Phys. Rev. D \textbf{68}, 046005 (2003),
arXiv:hep-th/0301240 [hep-th].









\end{thebibliography}
\end{document}